\documentclass[aps,pre,preprint,epsfig,superscriptaddress]{revtex4-1}
\usepackage{amsmath,amssymb,graphicx}
\usepackage{latexsym}
\usepackage{esint}
\usepackage{color}
\usepackage{float}
\usepackage{multirow}
\usepackage{hyperref}
\usepackage{ulem}

\usepackage{enumerate}

\usepackage{amsbsy}
\usepackage{bm}

\newcommand{\be}{\begin{equation}}
\newcommand{\ee}{\end{equation}}
\newcommand{\bea}{\begin{eqnarray}}
\newcommand{\eea}{\end{eqnarray}}

\newcommand{\comment}[1]{}

\begin{document}

\title{Nonequilibrium critical dynamics of the\\ two-dimensional $\pm J$ Ising model}

\author{Ramgopal Agrawal$^1$, Leticia F. Cugliandolo$^{1,2}$, Lara Faoro$^3$, Lev B. Ioffe$^3$, and Marco Picco$^1$\\
{\small $^1$Sorbonne Universit\'e, Laboratoire de Physique Th\'eorique et Hautes Energies}, \\
{\small CNRS UMR 7589, 4 Place Jussieu, 75252 Paris Cedex 05, France}
	\\
{\small $^2$Institut Universitaire de France, 1 rue Descartes, 75231 Paris Cedex 05, France}
	\\
{\small $^3$Google Research, Mountain View, CA, USA}
}
%\email{}
%\affiliation{.}
%\date{\today}

\begin{abstract}
The $\pm J$ Ising model is a simple frustrated spin model, where the exchange couplings independently take the discrete value $-J$ with probability $p$ and $+J$ with probability $1-p$. It is especially appealing due to 
its connection to quantum error correcting codes. Here, we investigate the nonequilibrium critical behavior of the two-dimensional $\pm J$ Ising model, after a quench from different initial conditions to a critical point $T_c(p)$ on the \textit{paramagnetic-ferromagnetic} (PF) transition line, especially, above, below and at the \textit{multicritical Nishimori point} (NP). The \textit{dynamical} critical exponent $z_c$ seems to exhibit non-universal behavior for quenches above and below the NP, which is identified as a pre-asymptotic feature due to the \textit{repulsive} fixed point at the NP. Whereas, for a quench directly to the NP, the dynamics reaches the asymptotic regime with $z_c \simeq 6.02(6)$. We also consider the geometrical spin clusters (of like spin signs) during the critical dynamics. Each universality class on the PF line is uniquely characterized by the stochastic Loewner evolution (SLE) with corresponding parameter $\kappa$. Moreover, for the critical quenches from the paramagnetic phase, the model, irrespective of the frustration, exhibits an emergent \textit{critical percolation} topology at the large length scales.
\end{abstract}

\maketitle
\thispagestyle{empty}

\section{Introduction and Background}
\label{S1}
\setcounter{page}{1}

Frustrated magnetic systems are ubiquitous in nature, with applications ranging from neural networks to quantum error correction codes~\cite{Edwards_1975,sourlas1989spin,doi:10.1142/2460,nishimori2001statistical,Kitaev_1997,KITAEV20032,doi:10.1063/1.1499754,berthier2011dynamical}. To understand their physics one simple and established pathway is to start with a model having the key ingredients of the system of interest. In this direction, the $\pm J$ Ising model~\cite{sourlas1989spin,nishimori2001statistical,Dotsenko_1982,PhysRevLett.61.625,PhysRevB.55.1025,PhysRevLett.87.047201,PhysRevB.65.054425} has been quite popular. This simple model has many rich features, e.g., different universality classes of second order phase transitions, emergence of a spin glass phase, nontrivial fixed points, etc. Its Hamiltonian is defined as
\be
\label{eq1}
H = - \sum_{\langle ij \rangle} J_{ij} S_i S_j
\; .
\ee
Here, $S_i = \pm 1$ are Ising spins, placed at each site $i$ of the lattice. The subscript $\langle ij \rangle$ denotes a sum over all nearest-neighbor pairs, and the exchange couplings $J_{ij}$ are quenched random variables, 
taking values $\pm J ~(J>0)$ from a bimodal distribution
\be
\label{eq2}
P(J_{ij}) = p \delta(J_{ij} + J) + (1-p) \delta(J_{ij} - J)
\; .
\ee
Clearly, the variable $p$ is a parameter which introduces frustration. The pure Ising model is recovered for $p=0$, while the bimodal Ising spin glass is obtained for $p=1/2$.

Notably, the model~\eqref{eq1} has a finite-temperature spin glass phase in spatial dimension $d > 2$. The two-dimensional ($d=2$) model exhibits spin glass ordering at temperature $T = 0$ only and for $p>p_0 \simeq 0.103$. Numerous studies~\cite{PhysRevLett.87.047201,PhysRevB.65.054425,Picco_2006,PhysRevB.73.064410,PhysRevE.79.021129,PhysRevE.77.051115,parisen2009strong,PhysRevE.78.011110} have shown various intriguing properties in two dimensions, due to which the model has also gained some attention from the fields outside the classical statistical mechanics; e.g., the $2d$ $\pm J$ Ising model plays an important role in determining the error correction threshold for a certain class of Toric codes~\cite{Kitaev_1997,KITAEV20032,doi:10.1063/1.1499754,WANG200331,PhysRevLett.103.090501,PhysRevLett.120.180501}. Therefore, this is the topic of the present paper. Before detailing the problem under consideration, let us first look into the background of the model.

\begin{figure}[t!]
	\centering
	\rotatebox{0}{\resizebox{.95\textwidth}{!}{\includegraphics{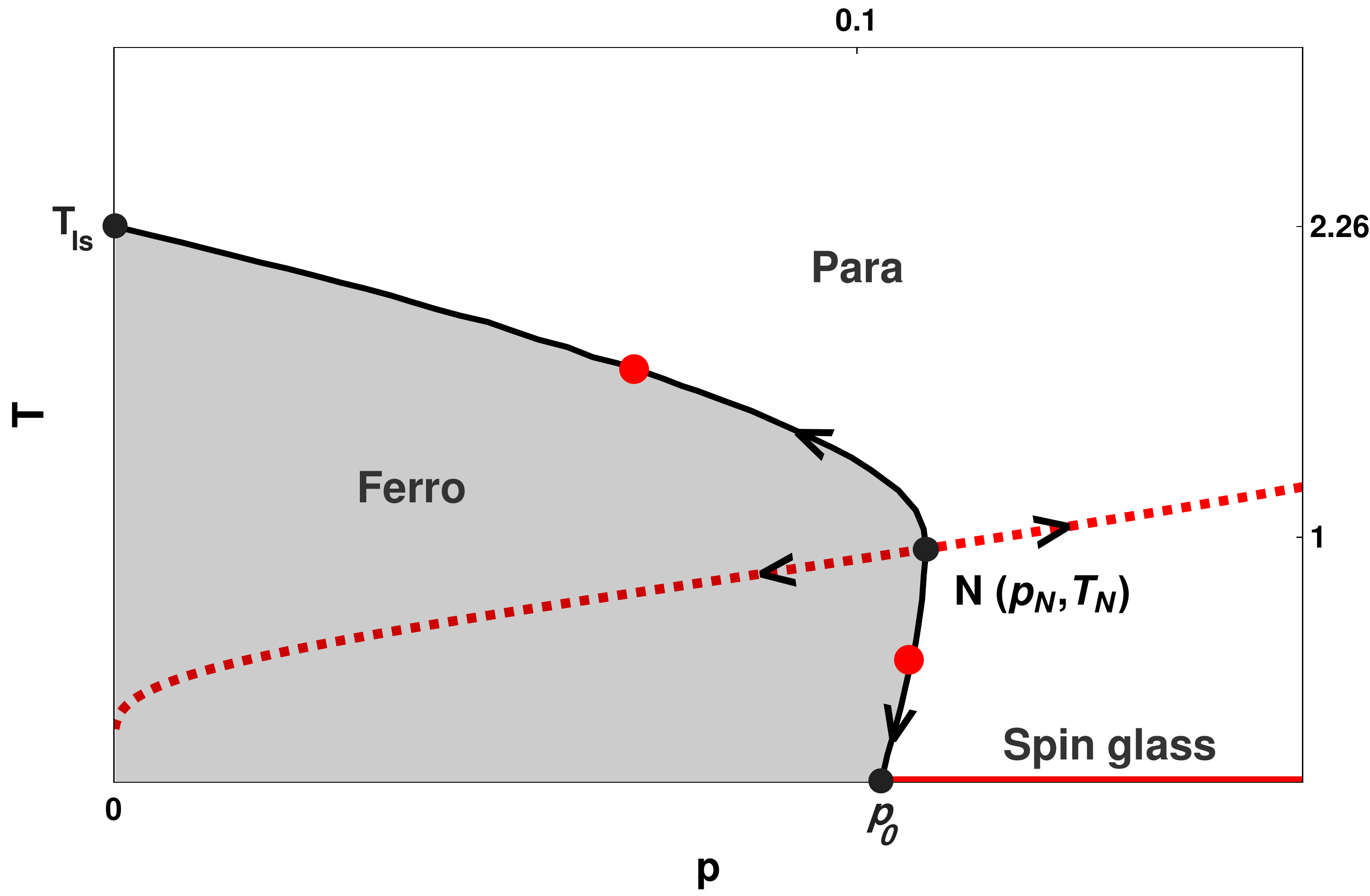}}}
	\vspace{-0.4cm}
	\caption{Phase diagram of the two-dimensional $\pm J$ Ising model. The black solid line locates the paramagnetic-ferromagnetic (PF) transition line, while the red dashed one represents the Nishimori line. The symbol $N$ denotes the Nishimori (multicritical) point. The arrows indicate the renormalization group (RG) flow off the Nishimori point. The filled circles on the PF line indicate the critical points where we quench the system during the simulations; the RG fixed points are particularly marked in black.}
	\label{fig1}
\end{figure}

The $p-T$ phase-diagram of the $2d$ $\pm J$ Ising model is shown in Fig.~\ref{fig1} (with $T$ measured in 
units of $J/k_B$). For small amount of disorder in terms of antiferromagnetic bonds, i.e, $0<p \lesssim p_0$, the model exhibits a paramagnetic-ferromagnetic (PF) phase boundary. Apart from that, due to a local gauge symmetry, there is
also a peculiar curve, known as the \textit{Nishimori line}, which is defined as~\cite{nishimori2001statistical,10.1143/PTP.66.1169}
\be
\label{eq3}
e^{-2 \beta J} = \frac{p}{1-p}
\; ,
\ee
where $\beta = 1/(k_B T)$ is the inverse-temperature and $k_B$ is Boltzmann constant. Some physical quantities, e.g., the internal energy, can be exactly calculated along this line. Most importantly, the Nishimori line is invariant under the renormalization group (RG) transformation, akin to the PF transition line. Therefore, the intersection point where the two lines meet is a multicritical fixed point, also known as the \textit{Nishimori point} (NP). Notice that the multicritical behavior at this special point was first pointed out by McMillan~\cite{PhysRevB.29.4026} in the $2d$ Ising model with Gaussian disorder.

The PF critical line starts from the Ising fixed point located at $T = T_{\rm Is}$ and $p=0$. With increase in $p$ it bends downwards and meets the Nishimori line at the point $(p_{\rm N},T_{\rm N})$. Notice that for $T > T_{\rm N}$ there is only one fixed point on the PF line, which is the Ising one. The disorder, for $T_{\rm N} < T < T_{\rm Is}$, is a \textit{marginally} irrelevant perturbation to the Ising fixed point~\cite{Dotsenko_1982,Picco_2006,PhysRevE.78.011110}. Therefore, the RG flow is attracted towards the Ising fixed point. Beyond the NP the PF line shows a re-entrant behavior and ends at another nontrivial fixed point located at $T=0$ and $p=p_0 < p_{\rm N}$, known as the \textit{strong disorder} fixed point~\cite{parisen2009strong,WANG200331,PhysRevB.70.134425}. As shown in Fig.~\ref{fig1}, this zero-temperature fixed point is a junction of ferromagnetic, paramagnetic, and spin glass phases. The latter exists at $p > p_0$ and $T=0$. In early studies~\cite{PhysRevB.54.364,PhysRevB.60.6740}, the universality class of this point was claimed to be that of critical percolation. However, it was later found~\cite{PhysRevLett.87.047201,PhysRevB.65.054425,PhysRevE.77.051115,parisen2009strong} with good numerical precision that the universality classes of both the NP as well as the strong-disorder fixed point are different from percolation and Ising classes. In fact, the critical exponents at these points clearly indicate the emergence of two new universality classes of second order phase transitions. Furthermore, the transition along the entire PF line for $T < T_{\rm N}$ is controlled by the strong-disorder fixed point~\cite{PhysRevB.29.4026,parisen2009strong}.

It is now clear that the $2d$ $\pm J$ Ising model possesses a quite rich critical behavior due to frustration. To further dig into the novel aspects of criticality, people have been interested in the non-equilibrium critical dynamics --- the post quench critical dynamics of an initially disordered/ordered system, from both theoretical~\cite{RevModPhys.49.435,tauber_2014,janssen1989new,PhysRevB.40.304,PhysRevLett.77.3704,doi:10.1142/S021797929800288X,Godreche_2002} as well as experimental~\cite{PhysRevB.46.3452,PhysRevB.34.3513}  points of view. During the relaxation to a critical equilibrium configuration, the regions of critical correlations (similar to those in equilibrium  at the target critical point) grow. As the system size diverges $L \rightarrow$ $\infty$, the characteristic relaxation time associated with this nonequilibrium process also diverges. The dynamical scaling symmetry also enters in the picture with a critical scaling relation $ C(r,t) = C_{\rm eq}(r) \overline{F}\left(r/\xi (t)\right)$. Here, $C(r,t)$ is the time-dependent \textit{spatial} correlation function (see its definition in Sec.~\ref{S2}), and $\xi (t)$ is the time-dependent correlation length which grows in a power-law fashion, $\xi (t) \sim t^{1/z_c}$, with $z_c$ a \textit{dynamical} critical exponent~\footnote{Apart from the static critical exponents the dynamical critical exponents are also equally important to characterize a critical phenomenon. Other nontrivial dynamical exponents are $\lambda_c$ (exponent associated with two-time correlations) and $\theta_c$ (persistence exponent).}.

In this paper, we thoroughly investigate the nonequilibrium critical dynamics of the $2d$ $\pm J$ Ising model with  single-spin-flip Monte Carlo simulations, where the system is quenched from an infinitely high temperature to the different critical points on the PF transition line including the multicritical points $T = T_{\rm N}$ and $T = T_c(p_0)$. In some cases we also explore the critical dynamics after a start from an ordered or critical initial state.

We are particularly interested in how the dynamical exponent $z_c$ changes with increase in the amount of disorder $p$. In the pure $2d$ Ising model $(p=0)$, the value of $z_c$ is $z_c \simeq 2.17$, and it has been confirmed via numerical simulations~\cite{PhysRevE.56.2310,PhysRevB.62.1089,Blanchard_2012,Ricateau_2018} as well as various analytical methods, e.g., the real space RG approach~\cite{PhysRevLett.41.128,PhysRevB.24.1419}, high temperature series expansion~\cite{PhysRevB.47.869}, the damage spreading technique~\cite{Poole_1990}, etc. In contrast, the critical dynamics in the current problem remain largely unexplored, apart from a few preliminary studies~\cite{Ozeki_2007,doi:10.1143/JPSJ.81.074602}.

A very efficient way to analyse criticality is to study the dynamic properties of various kinds of geometric structures. We focus on how the masses of different geometrical spin clusters (those of like spin signs) evolve during the dynamics. It is well known that at the critical point the equilibrium correlation length diverges and an infinitely large system possesses \textit{fractal} structures at all scales $r$ such that $r_0 < r < \infty$ ($r_0$ being the lattice spacing). During the nonequilibrium evolution towards the critical point, $\xi (t)$ is finite; however, the fractality of growing structures/interfaces is maintained~\cite{Blanchard_2012,Ricateau_2018} at scales $r < \xi (t)$, and is identical to that of an equilibrium macroscopic system at the target critical point. In this way, the critical dynamics also provides a clean demonstration of the equilibrium geometrical features. We emphasize that due to the quenched antiferromagnetic bonds in the frustrated system, the known cluster Monte Carlo algorithms~\cite{newman1999monte,PhysRevLett.62.361} no longer avoid the critical slowing down. That is why, the geometrical features of the frustrated critical systems including the spin glasses have always been mysterious. The nonequilibrium critical dynamics circumvents this problem by directly examining the geometrical features at the growing length scales.

Another interesting property of $2d$ Ising systems is the emergence of \textit{critical percolation} structures. It is by now well established~\cite{BlaCorCugPic14,AreBrayCugSic07,Blanchard_2017,Ricateau_2018} that soon after a quench from the paramagnetic phase to $T \le T_c$ the system reaches the critical point of $2d$ random site percolation. In fact, this phenomenon is proven to be quite general and robust against quenched disorder~\cite{CorCugInsPic17}, dilution~\cite{CorCugInsPic19}, long range interactions~\cite{PhysRevE.105.034131}, etc. Therefore, it will be worth  seeing what happens in the presence of frustration. Notice that similar to Ising criticality the geometrical features in critical percolation are also fractal, though of a different kind.

With the above questions in mind, we quantitatively examine the dynamical properties of geometrical features using the conformal invariance property~\cite{PhysRevLett.60.2343,schramm2000scaling,CARDY200581,Smirnov:2006cjh}, according to which, the interfaces of the geometrical clusters in the $2d$ critical systems can be described by the stochastic Loewner evolution (SLE) with a diffusion parameter $\kappa$. The value of $\kappa$ is unique for an universality class of the second order phase transition and therefore serves as a mathematical tool for the characterization of different universality classes. The fractal dimensions of the spin clusters can be directly calculated from this parameter $\kappa$. Further details are discussed in the subsequent Sections.

The key observations of our study are as follows. 
\begin{enumerate}
\item The dynamical critical exponent $z_c$ seems to exhibit non-universal behavior for a quench above ($T_{\rm N} < T_c(p) < T_{\rm Is}$) and below ($T_c(p_0) < T_c(p) < T_{\rm N}$) the NP. This, however, is identified as a pre-asymptotic feature due to the competition with the repulsive fixed point at NP. On the other hand, for a quench directly to the NP, the dynamics reaches the asymptotic regime with an exponent $z_c \simeq 6.0$ .

\item The diffusion parameter $\kappa$, obtained from small length scales ($r < \xi(t)$), uniquely characterizes each universality class. We measure three different values, depending on $T_c>T_{\rm N}$, $T_c=T_{\rm N}$, or $T_c<T_{\rm N}$, irrespective of the initial quenched state.

\item For all the critical quenches from the paramagnetic phase, the large scale ($r > \xi(t)$) topology belongs to the critical percolation.
\end{enumerate}

This paper is structured as follows. In Sec.~\ref{S2}, we detail the methodology and observable quantities. In Sec.~\ref{S3} and~\ref{S4}, we present our main results. Section~\ref{S3} discusses the dynamical critical exponent for different critical quenches, and Sec.~\ref{S4} details the dynamical properties of geometrical features during the time evolution. Finally, in Sec.~\ref{S5}, we summarize the results obtained so far in this paper and we discuss some open points indicating  possible future directions of research. Appendices~\ref{A1} and~\ref{A2} present some additional large-scale simulations for different critical quenches.

\section{Methodology and Observable quantities}
\label{S2}

We study the nonequilibrium critical dynamics of the $2d$ $\pm J$ Ising model on a square lattice with periodic boundary conditions (PBCs) in both $x$ and $y$ directions. At time $t=0$, the system is prepared by assigning random values $(\pm 1)$ to each Ising spin $S_i$, which is equivalent to an infinite temperature paramagnetic spin configuration. The model system is then quenched to different points $T_c(p)$ on the PF transition line including the multi-critical points. We choose $T_c \simeq 1.687$ at $p = 0.07$, $T_{\rm N} \simeq 0.952$ at $p \simeq 0.109$, $T_c \simeq 0.50$ at $p \simeq 0.107$, and $T_c=0$ at $p=p_0$, which are taken from Refs.~\cite{PhysRevE.77.051115,parisen2009strong}. For the time evolution of the spin configuration after a critical quench we exploit the Metropolis algorithm~\cite{metropolis1949monte,newman1999monte} with nonconserved order parameter kinetics. In this algorithm, a single spin flips with Metropolis transition rate
\be
W(S_i\to -S_i) = N^{-1}\min \left\{1,e^{-\frac{\Delta E}{T_c}}\right \}
\; ,
\label{metrop}
\ee
where $\Delta E$ is the energy difference in the proposed move and we have set to unity the Boltzmann constant. Time is measured in terms of Monte Carlo steps (MCS), each corresponding to $N = L^2$ attempted elementary moves. Notice that the configuration of bonds $\{J_{ij} \}$ is drawn from the probability distribution in Eq.~(\ref{eq2}) 
and is kept fixed during the time evolution.

One of the main observables we consider in our study is the time-dependent correlation length, which can be extracted from the spatial correlation function
\be
\label{corr}
C(r,t) = \langle  S_i(t) S_{i+\vec r}(t) \rangle - \langle  S_i(t) \rangle \langle  S_{i+\vec r}(t) \rangle 
\; ,
\ee
where $\langle \dots \rangle$ is a non-equilibrium average, taken over different random initial conditions and disorder realizations. For $r\gg r_0$ it obeys the following scaling relation during the dynamical scaling,
\be
\label{scale}
C(r,t) = \frac{1}{r^{\eta}} \overline{F}\left(\frac{r}{\xi (t)}\right) 
\; ,
\ee
where $\overline{F}(s)$ is a scaling function with $\overline{F}(0) = 1$. The correlation length $\xi (t)$ is defined as the average distance over which critical correlations have spread at time $t$. Clearly, $C(r,t)$ crosses over to the equilibrium correlation function $C_{\rm eq}(r)$ as $t \rightarrow \infty$,
\be
\label{corr_eq}
C_{\rm eq}(r) = \frac{{\rm e}^{-r/\xi_{\rm eq}}}{r^{\eta}}
\; ,
\ee
with $\xi (t) \rightarrow \xi_{\rm eq}$. Here, $\eta$ is the \textit{static} critical exponent.

We extract $\xi (t)$ from the fall of the function $F(r,t) = r^{\eta} C(r,t)$ as, $F(r=\xi (t),t) = F_0$. We fix the constant $F_0$ to $F_0 = 0.2$. This method is widely accepted to calculate $\xi (t)$ in the scaling regime~\cite{PhysRevE.103.012108}. In order to quantitatively examine the dynamical critical exponent $z_c$ from the asymptotic growth law, $\xi (t) \sim t^{1/z_c}$, we consider the effective exponent $z_{\rm eff}(t)$ defined as,
\be
\label{z_eff}
\frac{1}{z_{\rm eff}(t)} = \frac{d\ln \xi (t)}{d\ln t}
\; ,
\ee
and we study its behaviour at long times.

One can also estimate $z_c$ from the short-time-critical dynamics (STCD) approach~\cite{janssen1989new,doi:10.1142/S021797929800288X,Albano_2011}. The main idea is that the time evolution after the critical quench is also critical (on the scale of the time-dependent correlation length). Therefore, the characteristic features of the target critical point, e.g., scale invariance, should remain valid during the short-time dynamics as well. For the start from an ordered initial state, the following scaling Ansatz~\cite{janssen1989new,Albano_2011} for the $k^{\rm th}$ moment of the magnetization density was proposed,
\be
\label{scale_mk}
M_k(t,\tau,L) = \ell^{-k \beta/\nu} M_k(\ell^{-z_c}t,\ell^{1/\nu}\tau,\ell^{-1}L)
\; ,
\ee
where $\ell$ is a rescaling spatial parameter, $\tau$ is the reduced temperature, $L$ is the system size, $\beta$ and $\nu$ are the usual static critical exponents, and $z_c$ is the dynamical critical exponent. For large system sizes, the above expression predicts a simple power-law decay for the magnetization density,
\be
\label{m1}
M(t) \sim t^{- \beta / \nu z_c}
\; .
\ee
Later on, we will see the importance of the above decay law. Furthermore, the dynamical length scale and the associated exponent $z_c$ can be independently determined from the STCD by calculating the time-dependent Binder cumulant~\cite{Albano_2011,DASILVA2002325},
\be
\label{bind}
U(t,\tau,L) = \frac{M_2(t,\tau,L)}{\left[M(t,\tau,L)\right]^2} -1
\; ,
\ee
where $M_2(t,\tau,L)$ is the second moment of magnetization density $M(t,\tau,L)$. For large system sizes and quench to the critical point ($\tau_c$), this quantity shows power-law increase in time
\be
\label{bind1}
U(t) \sim t^{d/z_c}
\; .
\ee
Here $d$ is the dimensionality. One can see from the above relation that the quantity $\left[U(t)\right]^{1/d}$ also serves as a dynamical length scale in the system, which should be equivalent to the time-dependent correlation length $\xi (t)$ (see Appendix~\ref{A2}).

We investigate the geometry of growing domains/interfaces in the system by invoking some results from Conformal Field Theory (CFT). It has been found~\cite{PhysRevLett.60.2343,schramm2000scaling,CARDY200581} that under conformal invariance the interfaces (external hulls of clusters) of a $2d$ critical system can be described in terms of the SLE with diffusion parameter $\kappa$. This $\kappa$ controls the deviation of a curve from the straight line behavior and generates a unique conformal invariant curve for each $\kappa > 0$. Therefore, it uniquely characterizes the geometry of a critical system. It can also be related~\cite{PhysRevLett.84.1363,Blanchard_2012} to the Hausdorff dimensions by the following relations
\be
\label{frac_dim}
d_{{\rm l}} = 1 + \frac{\kappa}{8} 
\; , \qquad\qquad d_{{\rm A}} = 1 + \frac{3 \kappa}{32} + \frac{2}{\kappa}
\; ,
\ee
where $d_{{\rm l}}$ and $d_{{\rm A}}$ are the Hausdorff or fractal dimensions related to the interface and the area of the critical clusters, respectively.

In a lattice system, the above diffusion parameter $\kappa$ can be extracted from the (average) squared winding angle $\langle \theta^2(r) \rangle$. For a typical cluster this quantity is calculated as follows. At first, two  points, say $i,j$, are chosen at random at a distance $r$ on the external hull of the cluster. One then calculates the winding angle $\theta(r)$ between them by fixing a particular direction of rotation, say counterclockwise. Further, by averaging the square of $\theta(r)$ over all possible couples of hull points at distance $r$ (one can also include ensemble averages for better statistics), one finds the quantity $\langle \theta^2(r) \rangle$. For $2d$ critical systems, this quantity satisfies~\cite{PhysRevLett.60.2343,PhysRevE.68.056101}
\be
\label{wind}
\langle \theta^2(r) \rangle = a + b(k) \ln\left(\frac{r}{r_0}\right)
\; ,
\ee
where $a$ is a nonuniversal constant, and $r_0$ is the lattice spacing. The slope $b(k)$ is a function of the diffusion parameter $\kappa$:
\be
\label{eq:b-def}
b(k) = \dfrac{4\kappa }{ 8 + \kappa }
\; .
\ee
Notice that $\kappa = 2$ corresponds to the loop-erased random walks~\cite{10.2307/3481623}, $\kappa = 3$ to the interfaces of critical Ising model~\cite{smirnov2010conformal}, and $\kappa = 6$ to the critical percolation interfaces~\cite{SMIRNOV2001239}. For the fractal interfaces~\cite{PhysRevLett.60.2343} the winding angle $\theta(r)$ is Gaussian distributed with zero mean. That is why, the above quantity $\langle \theta^2(r) \rangle$ is also referred to as \textit{winding angle variance} (WAV).

Since we are interested in the nonequilibrium dynamics, the WAV depends on time, i.e., $ \langle \theta^2(r, t) \rangle$. In fact, as we will discuss in further Sections, the slope $b(\kappa)$ and so the value of the diffusion parameter $\kappa$ obtained from Eq.~\eqref{wind} vary with space $(r)$ and time $(t)$.

In the next Sections we present our numerical results for the $2d$ $\pm J$ Ising model. We consider square lattices of two different linear sizes, $L=128$, and $L=1024$. For $L=128$ the numerical data are averaged over at least $2000$ (sometimes more) runs, while for $L=1024$ the data are averaged over $1000$ runs, with each run consisting of different initial configuration of spins $\{S_i(0)\}$ and disorder realization $\{J_{ij} \}$.

\section{Dynamical critical exponent}
\label{S3}

\begin{figure}[t!]
	\centering
	\rotatebox{0}{\resizebox{.85\textwidth}{!}{\includegraphics{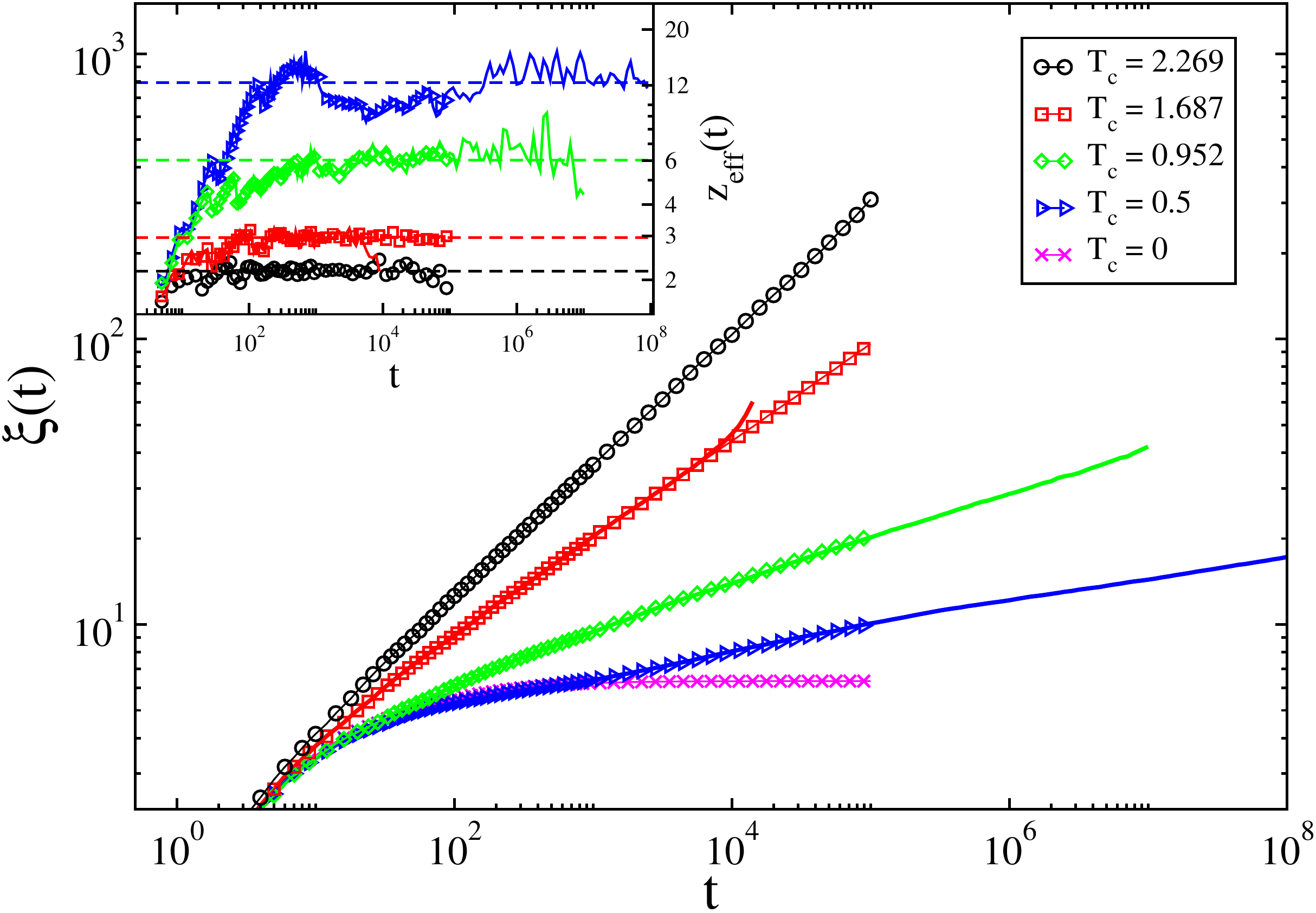}}}
	\caption{Plot of the correlation length $\xi (t)$ vs time $t$, in log-log scale, for different critical quenches (see the key) from an infinitely high temperature $T \gg T_c$. The symbols represent data for a system of linear size $L = 1024$, while the solid lines correspond to those of $L = 128$. The inset plots the effective exponent $z_{\rm eff}$ vs $t$ in log-log scale for the datasets in the main frame. The dashed horizontal lines indicate the late-time plateau of $z_{\rm eff}$.}
	\label{fig2}
\end{figure}

Let us start by discussing the growth law of the time-dependent correlation length. In Fig.~\ref{fig2}, the correlation length $\xi (t)$ is plotted against time $t$ for different critical quenches from the paramagnetic phase on the PF transition line. In case in which the dynamics is too slow, we also analyzed the growth law using a smaller system $(L=128)$ apart from the larger one  $(L=1024)$. Notice that for smaller sizes the simulations are quick and the asymptotic regime becomes accessible, while for larger sizes the results are free from finite-size-effects. In the inset the effective exponent $z_{\rm eff}$ (\ref{z_eff}) is plotted against $t$. Its late time saturated value provides the dynamical critical exponent $z_c$. 

When disorder $p$ is zero, i.e., at $T_c = T_{\rm Is}$, the asymptotic growth law is $\xi (t) \sim t^{1/z_c}$, which is clearly observed in the log-log plot of Fig.~\ref{fig2}, and the value of the dynamical critical exponent $z_c$ is in excellent agreement with the known estimate $z_c \simeq 2.17$ (see Tab.~\ref{tab1}).

\begin{figure}[t!]
	\centering
	\rotatebox{0}{\resizebox{.65\textwidth}{!}{\includegraphics{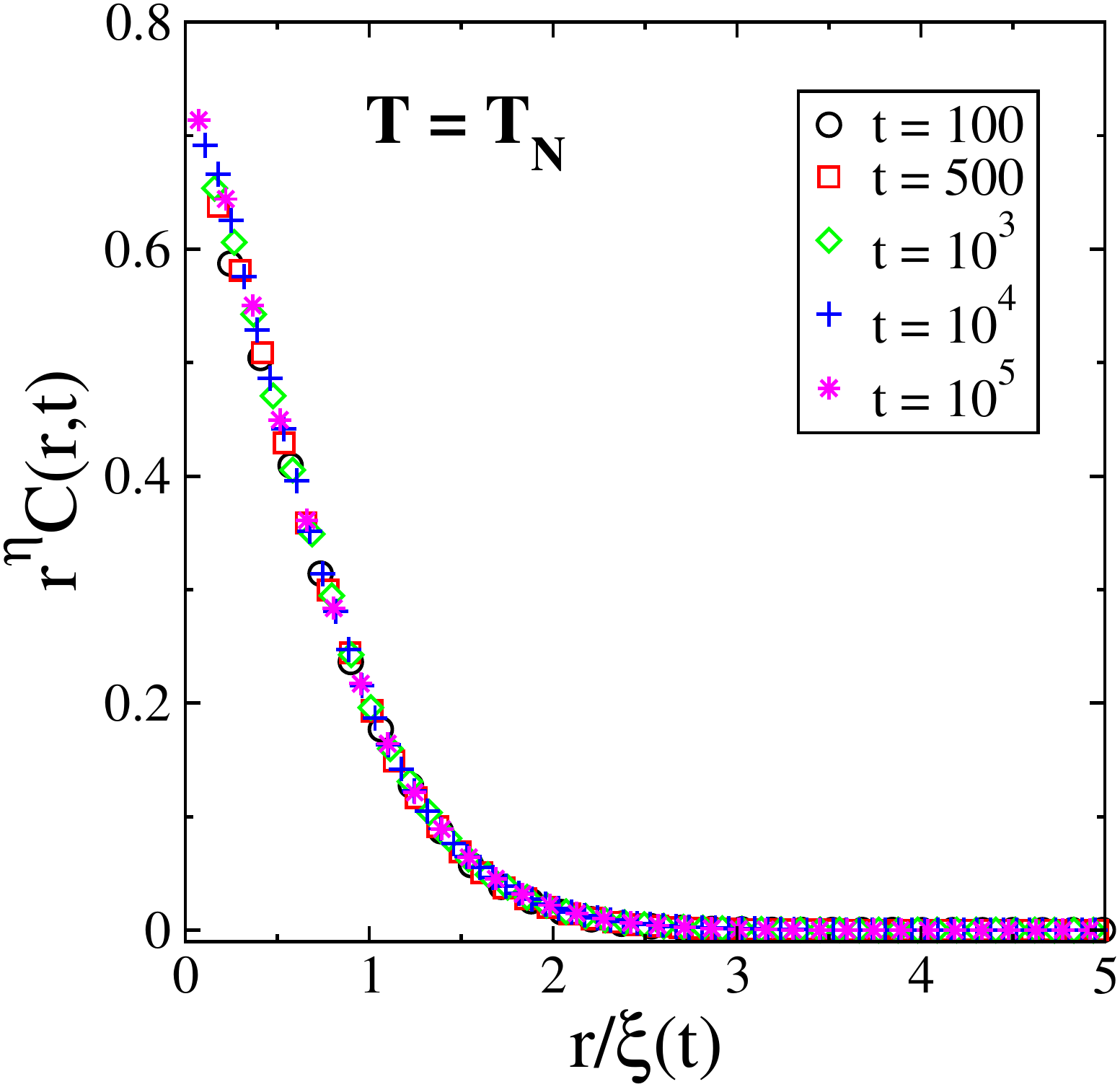}}}
	\caption{Plot of $r^{\eta} \, C(r,t)$ against the scaling variable $r / \xi(t)$ for a quench from infinitely high temperature $T \gg T_{\rm N}$ to  $T_{\rm N}$ of a system of linear size $L = 1024$. Different datasets represent different timesteps (see the key). The parameter  $\eta ~(\simeq 0.177)$ is the static critical exponent for the equilibrium correlation function (see the main text).}
	\label{fig3}
\end{figure}

We next study the disordered case, i.e., $p \ne 0$. We first discuss the law for a quench to the Nishimori point $(T_{\rm N})$. The inset in Fig.~\ref{fig2} shows that for the long timescales the effective exponent $z_{\rm eff}$ saturates to a constant value, $z_c \simeq 6.0$. To crosscheck this observation, we repeat the same simulations from an ordered initial state and also till much larger timescales (not shown here). We find that the exponent $z_c$ remains unchanged. This concludes that the current value $z_c \simeq 6.0$ is in the asymptotic regime. Notice that the value of the exponent $z_c$ is quite larger than that at the Ising fixed point, indicating the slow relaxation of the system at the NP. The dynamical scaling during the growth of critical correlations in terms of $\xi(t)$ can be observed in Fig.~\ref{fig3}.

Now we look at critical quenches between $T_{\rm Is}$ and $T_{\rm N}$. For a quench to $T_c \simeq 1.687$ at $p = 0.07$, the numerical data in Fig.~\ref{fig2} support an exponent around $z_c \simeq 2.95$ for both system sizes $L=128$ and $L=1024$. At first glance, this observation seems in contrast to the universality predictions~\cite{Dotsenko_1982,Picco_2006,PhysRevE.78.011110}, according to which, the dynamic critical exponent $z_c$ should \textit{asymptotically} tend to the Ising value $(\simeq 2.17)$ for all quenches to $T_c(p) > T_{\rm N}$ on the PF line. We remind that the disorder on the PF line above NP is just a marginally irrelevant perturbation. Therefore,  this large value of $z_c$ is likely due to a preasymptotic behavior, which is explained as follows. The quench has been done to $T_c \simeq 1.687$ which lies between the attractive Ising and the repulsive NP fixed points. The competition between these fixed points will certainly ensue crossover effects also in the dynamics (see, e.g., Refs.~\cite{Janssen_1995,Heuer_1993}). For a quench to the NP we measured above an asymptotic value $z_c \simeq 6.0$. Likely, the approach to a constant value around $2.95$ in the current simulations is an effect of the fixed point at NP, and the true asymptotic exponent ($z_c \simeq 2.17$) should appear on still longer timescales and larger system sizes (see Appendix~\ref{A1}). We prompt the reader that due to disorder the precise characterization of the crossover might be hard and even unreachable, and that is why a preasymptotic regime is often misunderstood as a nonuniversal behavior~\cite{Silva_2009,Ozeki_2007,doi:10.1143/JPSJ.81.074602}. In Appendix~\ref{A1}, we have attempted hard to see some signatures of the crossover with large system sizes and the STCD method discussed in Sec.~\ref{S2}. Interestingly, we observed further decay in the dynamical exponent beyond the preasymptotic regime. In addition, the preasymptotic value decreases and the crossover time also shrinks as the disorder value $p$ is shifted towards the Ising point $T_{\rm Is}$. This explains the competition with the repulsive fixed point at NP. We also mention that in contrast to the dynamical critical exponents, the \textit{effective} static exponents soon reach their universal values and do not show any crossover behavior~\cite{Silva_2009,Ozeki_2007,doi:10.1143/JPSJ.81.074602}. The latter is expected as the static exponents are not related with the dynamics of the model.

Before considering a quench to a critical point between the NP and the strong disorder fixed point at $p=p_0$, let us first understand the critical dynamics right at the fixed point $(p=p_0,T_c = 0)$. In Fig.~\ref{fig2}, the plot of the correlation length $\xi (t)$ shows that after a time $t \sim 10^3$ from a quench at $t=0$ the growth is \textit{almost} frozen and the system is stuck in a metastable state. This is obvious due to the absence of thermal fluctuations, and indicates that the dynamical critical exponent $z_c$ is in practice divergent, i.e., $z_c = \infty$. With this information, we now proceed to an interesting case where $T_c \ne 0 < T_{\rm N}$. We choose $T_c \simeq 0.5$, with $p$ fixed to $p \simeq 0.107$. For this case, the effective exponent $z_{\rm eff}$ exhibits multi-plateau regimes beyond $t \sim 500$. The $z_{\rm eff}$ first stays around $z_c \simeq 10$. However, for late times ($t \sim 10^6$ onwards), it slightly shifts to a larger value $z_c \simeq 12$. Again, our understanding is that this strange behavior is due to the different fixed point at NP. The RG flow on the PF line below NP is attracted towards the strong disorder fixed point~\cite{parisen2009strong,PhysRevB.29.4026}. Therefore, the value of $z_{\rm eff}$ will increase indefinitely. However, given the slowness of the dynamics, observing a crossover to the true asymptotic regime in a real-time computation is far more challenging. Notice also that the flow will reach the fixed point only in the infinite size limit, i.e., $\lim_{L\rightarrow \infty} z_c = \infty$.

In Appendix~\ref{A2}, we have also extracted the effective exponent $z^{'}_{\rm eff}(t)$ [see Eq.~\eqref{z_eff1}] from the quantity $\left[U(t)\right]^{1/2}$ for different critical quenches above, below, and at $T_{\rm N}$ from the completely ordered state. We recall that this quantity measures the exponent $z_c$ independent of any static critical exponent, while the correlation length $\xi(t)$ inherently incorporates the universal value of critical exponent $\eta$ [see Eq.~\eqref{scale}]. We observe that the exponents $z^{'}_{\rm eff}(t)$ obtained from $\left[U(t)\right]^{1/2}$ are in good agreement with the $z_{\rm eff}(t)$ obtained from $\xi(t)$, indicating the similar values of $z_c$.

\begin{table}[t!]
	\begin{center}
		\begin{tabular}{| c | c | c | c | c | c |} 
			\hline 
			$T_c$ & $p$ & \text{universality class} & $d_{\rm l}$ & $\kappa$ & $z_c$ [from $\xi(t)$] \\
			\hline
			2.269 & 0 & \text{Ising} & 1.374 (1) & 2.99 (1)   & 2.17 (1) \\
			\hline
			1.687 & 0.07 & \text{Ising} & 1.372 (2) & 2.975 (9) & 2.95 (1) \\
			\hline
			0.952 & 0.109 & \text{NP} & 1.277 (2) & 2.22 (2) &  6.02 (6) \\
			\hline
			0.5 & 0.107 & \text{strong disorder} & 1.24 & 1.932 (4) & 12.3 (2) \\
			\hline
			0.0 & 0.103 & \text{strong disorder} & -- & -- & $\infty$ \\
			\hline
		\end{tabular}
	\end{center}
	\caption{The $2d$ $\pm J$ Ising model: dynamical critical exponent $z_c$, fractal dimension $d_{\rm l}$, and SLE diffusion parameter $\kappa$ for critical temperatures $T_c$ lying in different universality classes at paramagnetic-ferromagnetic (PF) line. The values are estimated from nonequilibrium dynamics after a quench from high temperature $T \gg T_c$ to $T_c$, where disorder parameter $p$ is fixed. For quenches between the two fixed points at PF line, $z_c$ denotes the preasymptotic value of the effective exponent.}
	\label{tab1}
\end{table}

\section{Dynamical properties of geometrical features}
\label{S4}

\subsection{Quenches to $T_c(p) > T_{\rm N}$}

It is well known that in equilibrium at the Ising critical point $T_{\rm Is}$ the system has geometric structures with (interfacial) fractal dimension $d_{\rm l} = 11/8$ (diffusion parameter $\kappa = 3$)~\cite{Smirnov:2006cjh,smirnov2010conformal}. Therefore, when a $2d$ Ising model is suddenly quenched from the paramagnetic phase to $T = T_{\rm Is}$, the Ising like fractality should hold at growing length scales $r < \xi(t)$, and that of the critical percolation (a feature of quenching from high temperature phase) should arise at length scales $r > \xi(t)$~\cite{AreBrayCugSic07,Blanchard_2012}. Here we particularly investigate the geometrical features in the presence of frustration, especially when the system is quenched to the marginally irrelevant perturbation regime of the PF line.

\begin{figure}[t!]
	\vspace{0.5cm}
	\centering
	\rotatebox{0}{\resizebox{.45\textwidth}{!}{\includegraphics{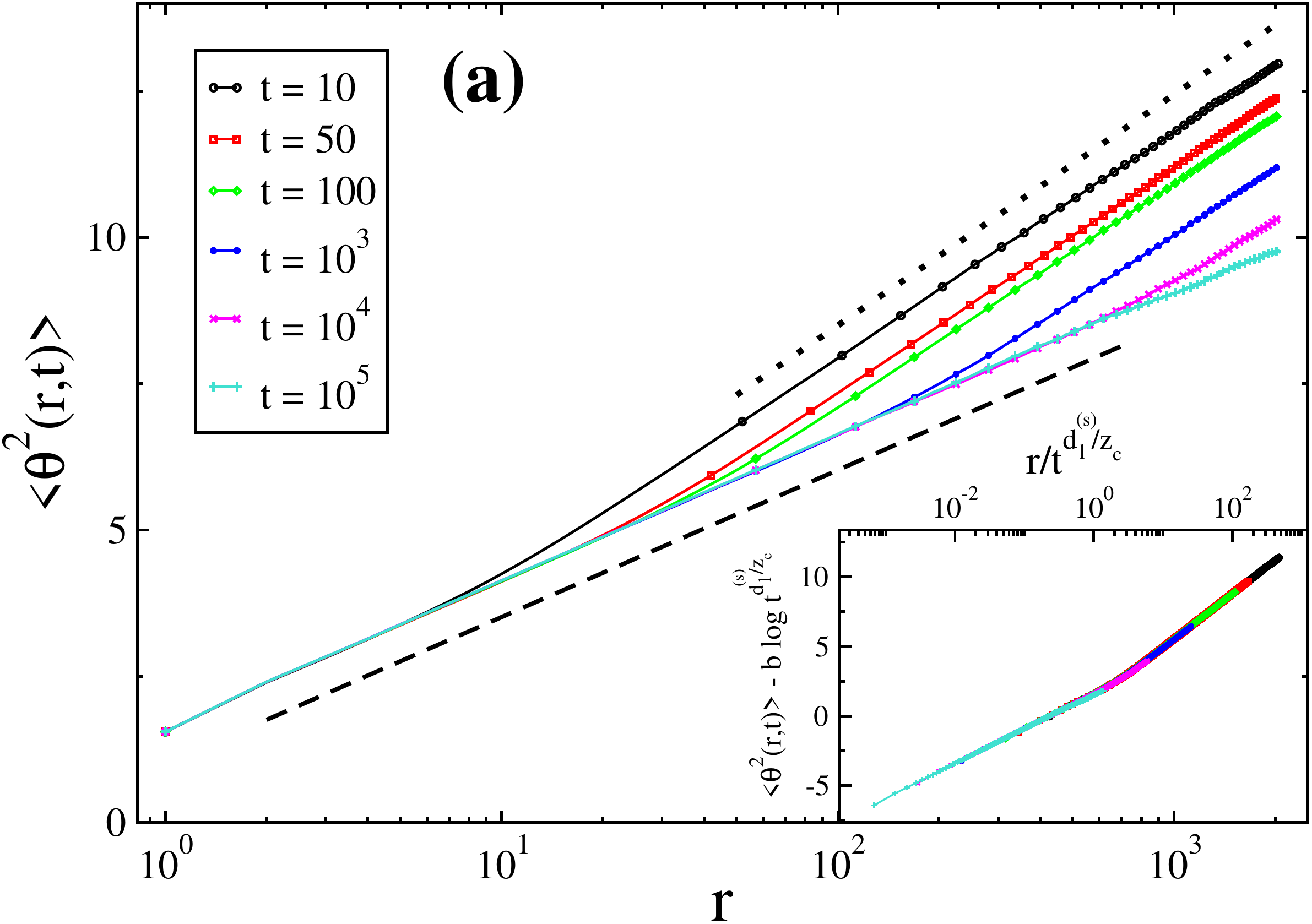}}}
	\rotatebox{0}{\resizebox{.45\textwidth}{!}{\includegraphics{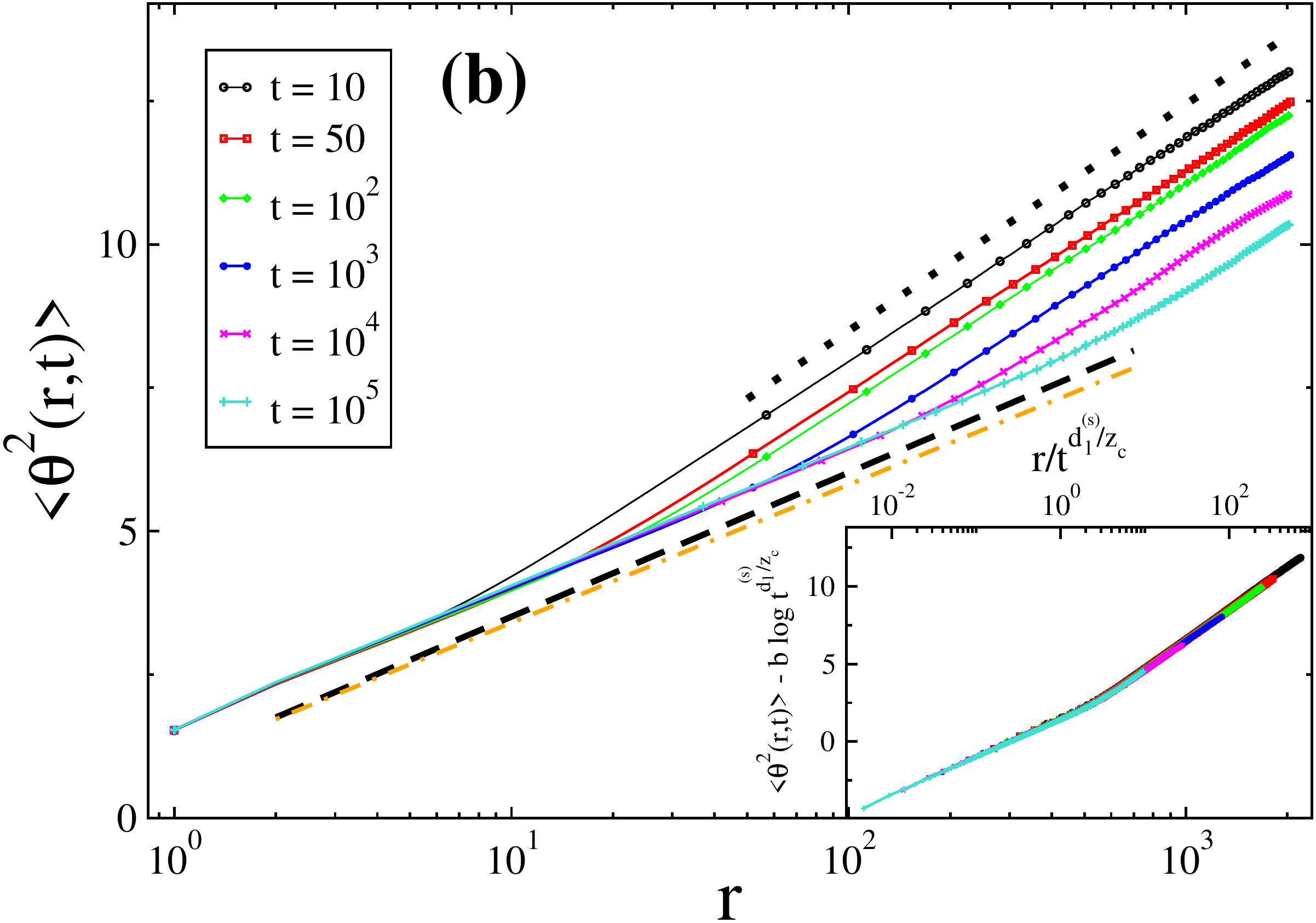}}}
	\caption{Winding angle variance (WAV) vs curvilinear distance $r$ for quenches from infinitely high temperature to (a) $T_{\rm Is}$ at $p=0$, and (b) $T_c = 1.687$ at $p=0.07$, in a system with linear size $L = 1024$. Different datasets are taken at different times after the quench (see the key). The dashed and dotted lines in both panels correspond to the stochastic Loewner evolution (SLE) with $\kappa = 3$ and $\kappa = 6$, respectively. The dot-dashed line in (b) represents SLE with $\kappa = 2.83$. The insets in (a) and (b) plot the quantity $\langle \theta^2(r, t) \rangle - b(\kappa^{(s)}) \ln t^{d_l^{(s)}/z_c}$ against $r / t^{d_l^{(s)}/z_c}$ for data in the respective main frames (see the main text for details).}
	\label{fig4}
\end{figure}

Let us first benchmark the disorder-free case~\cite{Blanchard_2012,Ricateau_2018}, i.e, a quench from the paramagnetic phase to $T = T_{\rm Is}$. In Fig.~\ref{fig4}(a), the behavior of the WAV $\langle \theta^2(r, t) \rangle$ is explored at different times. One can see that, up to a certain value of $r$ that increases with time, the slope of the curves at different times is similar to that of SLE with $\kappa=3$. Moreover, for large value of $r$, the SLE with $\kappa = 6$ is recovered, which belongs to the fractal structures at critical random percolation. Such a behavior is observed, because, a \textit{stable} critical percolation structure is formed at a time $t_p$~\footnote{The pinning time $t_p$ is much smaller than the relaxation time $t_{\rm eq} \simeq L^{z_c}$, and in the thermodynamic limit, $t_p/t_{\rm eq} \rightarrow 0$. See Ref.~\cite{BlaCorCugPic14} for details.} after the quench of the system from a high $T$ state at time $t=0$. This implies that the interfaces are fractal on all scales but with different fractal dimensions. Therefore, the crossover lengthscale, $r_{\rm cross} \propto [\xi(t)]^{d_l^{(s)}} \sim t^{d_l^{(s)}/z_c}$, where $d_l^{(s)}$ is the interfacial fractal dimension at small scales. One can write~\cite{CorCugInsPic19},
\be
\label{cross}
\langle \theta^2(r,t) \rangle - b(\kappa^{(s)}) \ln\left( t^{d_l^{(s)}/z_c} \right) = f\left( \frac{r}{t^{d_l^{(s)}/z_c}} \right),
\ee
where $\kappa^{(s)}$ denotes the diffusion constant (of SLE) related to the interfaces at small scale, i.e., $\kappa^{(s)} = 3$ in the present case, and $z_c$ is the dynamical critical exponent discussed above. The scaling function $f$ has the following limiting forms:
\be
\label{cross_r}
f(x) \sim \left\{
\begin{array}{lr}
	b(\kappa=\kappa^{(s)}) \ln x,  & ~~x \ll 1,
	\\
	\\
	b(\kappa=6) \ln x,  & ~~x \gg 1.
\end{array}
\right.
\ee
The parameter $b$ is defined in Eq.~(\ref{eq:b-def}). The perfect collapse of data in the inset of Fig.~\ref{fig4}(a) confirms the relations~\eqref{cross}-\eqref{cross_r}.

In Fig.~\ref{fig4}(b), the WAV is plotted for a quench from high $T \gg T_c$ to $T_c \simeq 1.687$, where the disorder parameter $p$ is fixed to $p = 0.07$. This value of critical point lies below the $T_{\rm Is}$ but above the $T_{\rm N}$ (see the phase diagram in Fig.~\ref{fig1}). Similarly to the disorder-free case, the behavior of the curves at small length scales which increases in time seems compatible with SLE with $\kappa=3$. When fitting the latest time curve ($t=10^5$) in a spatial window of $r\in [5,200]$, we find $\kappa \simeq 2.83$ (shown by dot-dashed line). However, when the fitting window is varied up to $r \sim 650 $, the value of $\kappa$ also slightly changes. We obtain $\kappa \simeq 2.98$ by averaging over various such windows (see Tab.~\ref{tab1}). It indicates that the fractality of the domains on the PF line above NP is similar to the one of the Ising criticality class. This was expected as the disorder in this region is a marginally irrelevant perturbation. Further, the large scale critical percolation features also persist, as the curves at large $r$ still follow the SLE with $\kappa=6$. The latter holds for curves at early times ($t \sim 10$) as well, which tells that a pinning time $t_p$ of stable critical percolation structure also exists in the presence of frustration. The crossover relation~\eqref{cross} is also justified in the inset.

\subsection{Quenches to $T_{\rm N}$}

We have seen above that a small amount of frustration in terms of antiferromagnetic bonds induces many new characteristics in the $2d$ $\pm J$ Ising model. The Nishimori point at $T_{\rm N}$ (see phase diagram in Fig.~\ref{fig1}) is one of them, which bifurcates the PF line into two different universality classes of second order phase transition governed by Ising and strong disorder fixed points, respectively. The nature of the phase transition across the NP is also second order with a unique universality class. Since the amount of disorder at the NP is tiny ($p_{\rm N} \simeq 0.109$), it is expected that the conformal invariance of the system continues to hold~\cite{PhysRevLett.87.047201}. However, since the universality class at the NP is different, the diffusion parameter $\kappa$ and so the fractal dimension $d_{\rm l}$ should be different from the Ising ones. In this subsection we quantitatively explore these features via quenches to the NP from different initial states.

\begin{figure}[t!]
\vspace{0.5cm}
	\centering
	\rotatebox{0}{\resizebox{.45\textwidth}{!}{\includegraphics{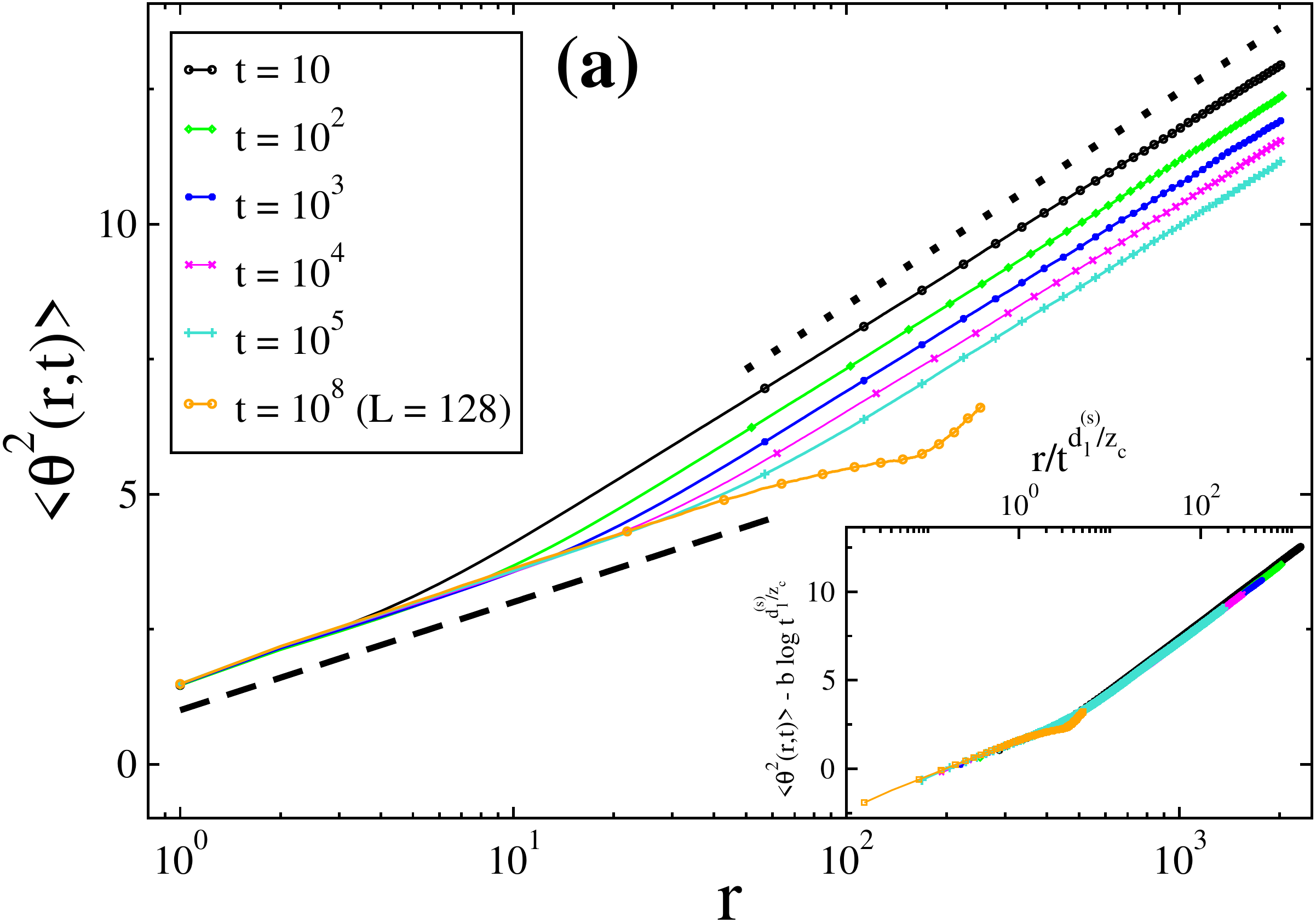}}}
	\rotatebox{0}{\resizebox{.45\textwidth}{!}{\includegraphics{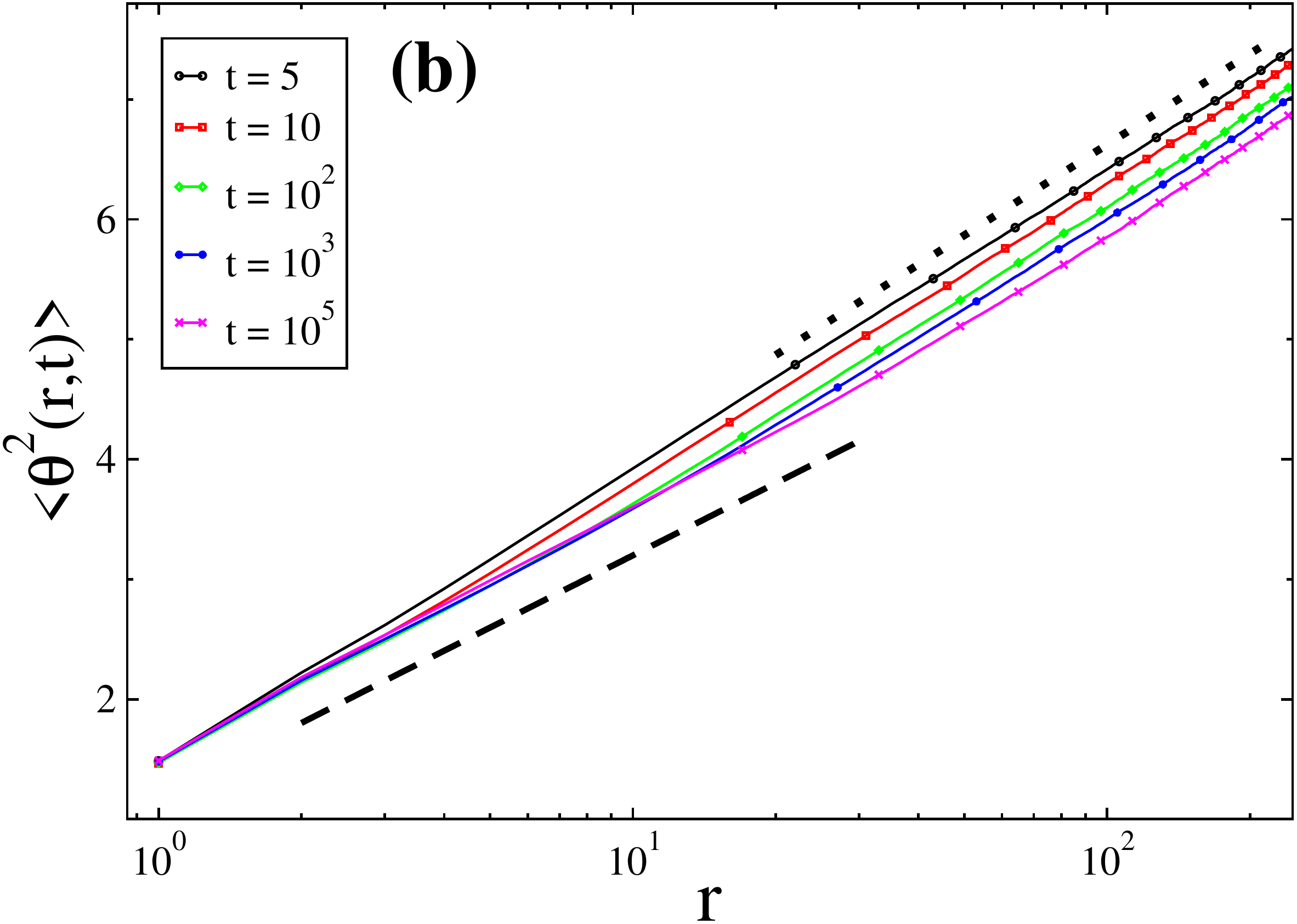}}}
	\caption{The winding angle variance (WAV) for quenches from (a) $T \gg T_{\rm N}$ and (b) $T_{\rm Is}$, both to $T_{\rm N}$. The system size for datasets at different times (see the keys) is $L = 1024$ unless mentioned explicitly. The dashed lines in both panels represent the stochastic Loewner evolution (SLE) with $\kappa \simeq 2.22$. The dotted lines represent the SLE with $\kappa = 6$ and $\kappa = 3$ in (a) and (b), respectively. The inset in (a) plots the quantity $\langle \theta^2(r, t) \rangle - b(\kappa^{(s)}) \ln t^{d_l^{(s)}/z_c}$ vs $r / t^{d_l^{(s)}/z_c}$ for data in the main frame (see the main text for details).}
	\label{fig5}
\end{figure}

In Fig.~\ref{fig5}(a) the WAV is shown at different times after a quench from infinitely high $T$ to $T_{\rm N}$. The slope of different curves up to a time-dependent value of length $r$ clearly indicates that the fractality of geometric features at NP is quite different from the one on the Ising point --- the slope at different times rather favors the SLE with $\kappa \simeq 2.22 ~(d_{\rm l} \simeq 1.27)$. This reconfirms the different universality class at the NP. Notice that due to the slow growth (large dynamical exponent) at the NP the WAV curve for system size $L = 1024$ and time $t=10^5$ shows compatibility with $\kappa \simeq 2.22$ till $r \simeq 10$. However, for a system of size $L=128$ and time $t=10^8$, the same slope prevails till $r \simeq 60$.

\begin{figure}[t!]
	\centering
	\rotatebox{0}{\resizebox{.7\textwidth}{!}{\includegraphics{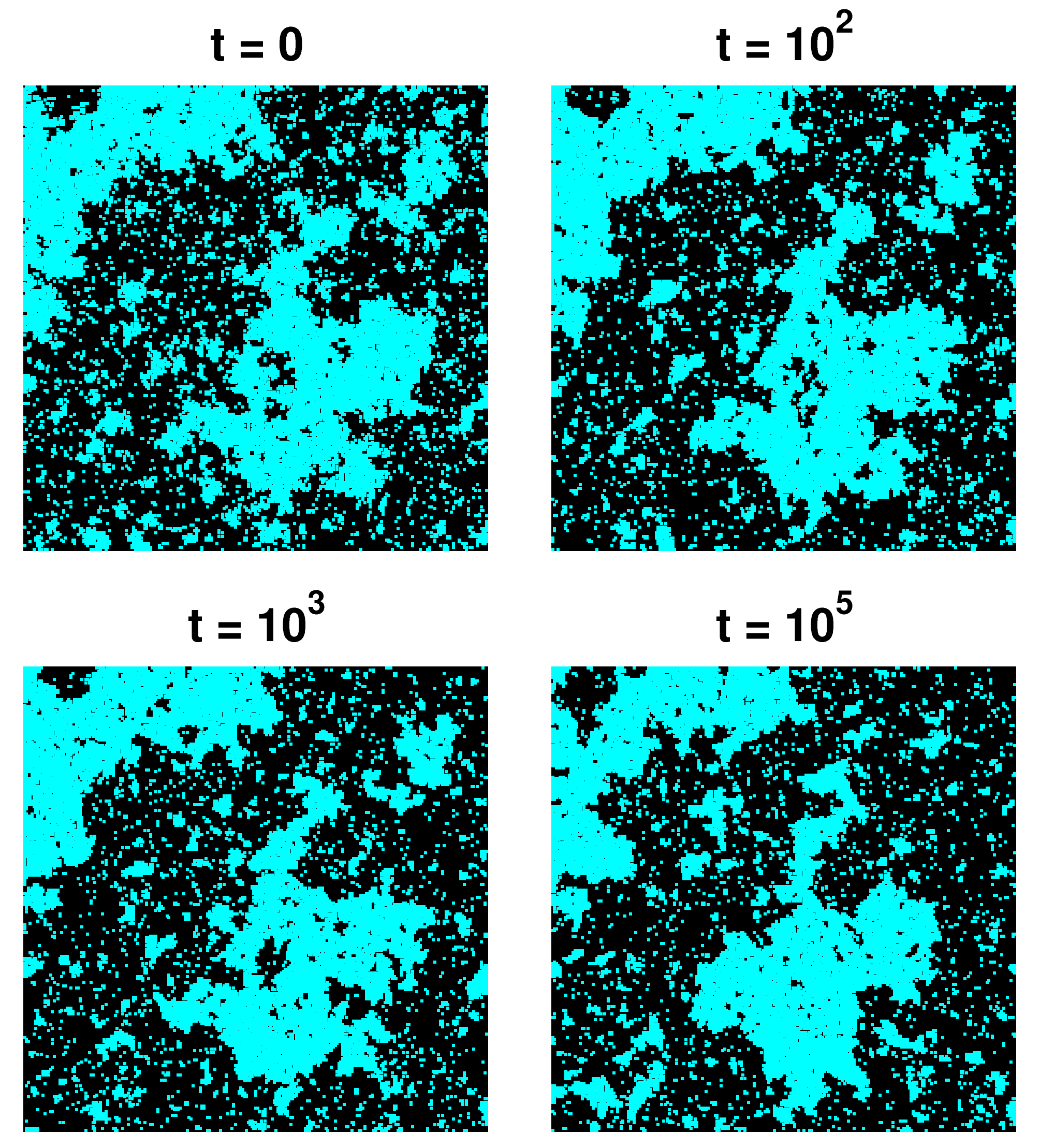}}}
	\caption{Instantaneous configurations of a system of linear size $L = 1024$ (we show only a $300^2$ portion of the full system), at different times given above the snapshots, 
	after a quench from $T = T_{\rm Is}$ to $T = T_{\rm N}$.}
	\label{fig6}
\end{figure}

For larger $r$, the slope of the WAV curves still follows the critical percolation behavior ($\kappa=6$), as shown in Fig.~\ref{fig5}(a). Further, the crossover between these two distinct behaviors can be checked by the relation~\eqref{cross}, by substituting $\kappa^{(s)} \simeq 2.22$ and $z_c \simeq 6.0$. In the inset of Fig.~\ref{fig5}(a) we plot the scaling variable $\langle \theta^2(r, t) \rangle - b(\kappa^{(s)}) \ln t^{d_l^{(s)}/z_c}$ against $r / t^{d_l^{(s)}/z_c}$ for data in the main frame. The nice collapse upholds the validity of the crossover~\eqref{cross}.

To rigor our understanding, we also quenched the system on $T = T_{\rm N}$ from a critical Ising state at $T = T_{\rm Is}$. We first prepared the initial spin configurations at $T_{\rm Is}$ using the Wolff cluster algorithm~\cite{newman1999monte,PhysRevLett.62.361}. The Metropolis algorithm~\eqref{metrop} was then exploited to evolve the system from $T_{\rm Is}$ to $T_{\rm N}$. Notice that contrary to a paramagnetic state the system at $T_{\rm Is}$ is power-law correlated,
\be
\langle S_i(t=0) S_{i+\vec{r}}(t=0) \rangle \propto \frac{1}{r^{\eta}},
\label{ising}
\ee
where $\eta = 1/4$. The system at $t=0$ already has fractal structure with $\kappa = 3$ (see the discussion above). Therefore, after quenching it to some other $T$, the critical percolation structures would not emerge. Rather, at large scales ($r > \xi(t)$), the system should have the fractality of the initial (Ising) class~\cite{Blanchard_2012}. This scenario is explained quite clearly in Fig.~\ref{fig5}(b). At small length scales the system has geometrical features with $\kappa \simeq 2.22$, while at large length scales the features of the Ising universality class persist. The evolution snapshots of the system after a quench from  $T_{\rm Is}$ to $T_{\rm N}$ are shown in Fig.~\ref{fig6}.

We finally conclude that the geometrical features at NP are described by SLE with $\kappa \simeq 2.22$, i.e., $d_{\rm l} \simeq 1.27$.

\subsection{Quenches to $T_c(p) < T_{\rm N}$}

Let us finally discuss the dynamical properties of the geometrical features after a quench from high $T \gg T_c$ to $T_c < T_{\rm N}$. As we discussed earlier, the critical behavior on this segment of the line is governed by the strong disorder fixed point at $T_c(p_0)$, which is zero. Further, the universality class of the transition is different from Ising and NP~\cite{parisen2009strong}. Therefore, it is expected that the fractal dimension may also be unique.

\begin{figure}[h!]
\vspace{0.5cm}
	\centering
	\rotatebox{0}{\resizebox{.45\textwidth}{!}{\includegraphics{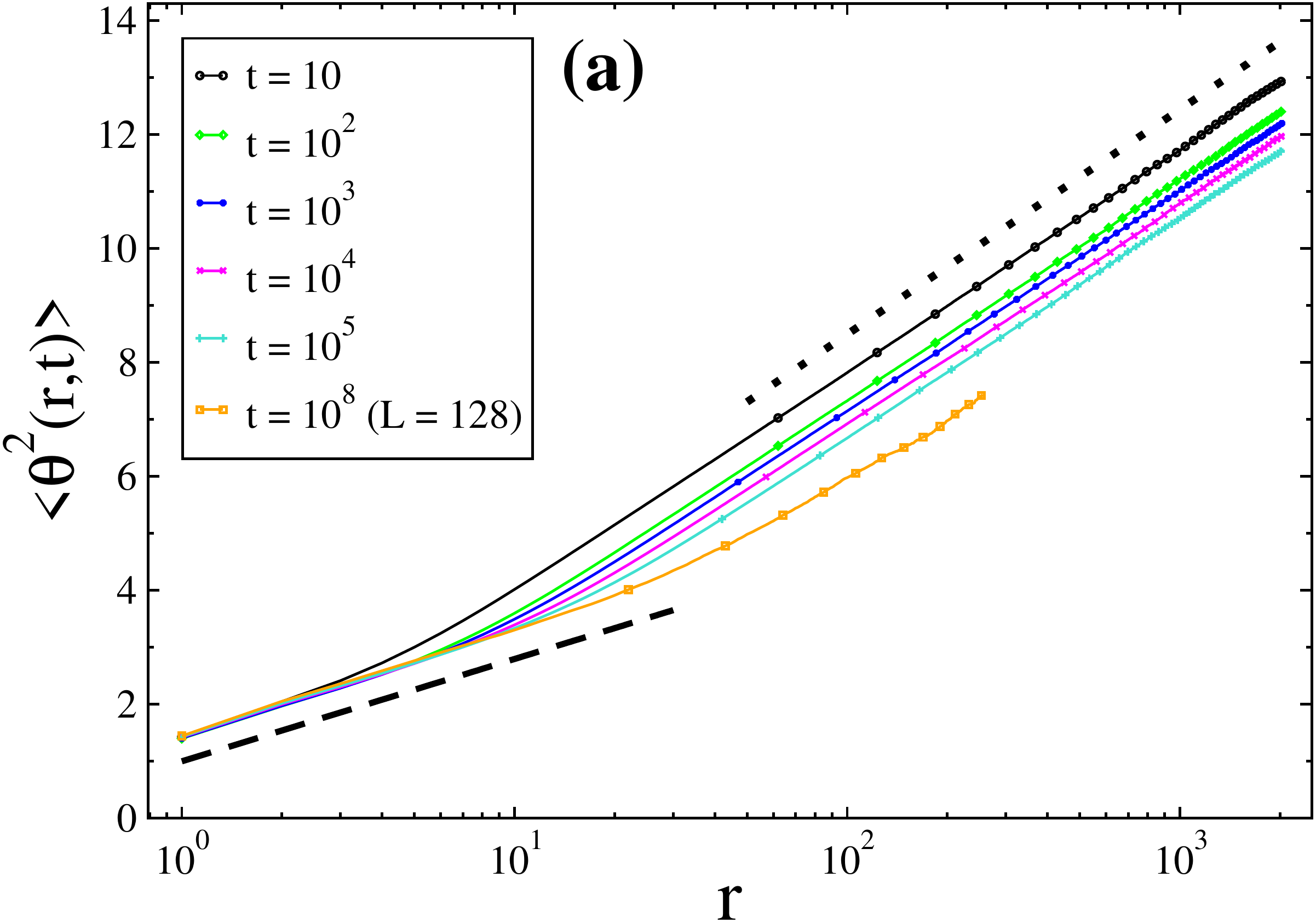}}}
	\rotatebox{0}{\resizebox{.45\textwidth}{!}{\includegraphics{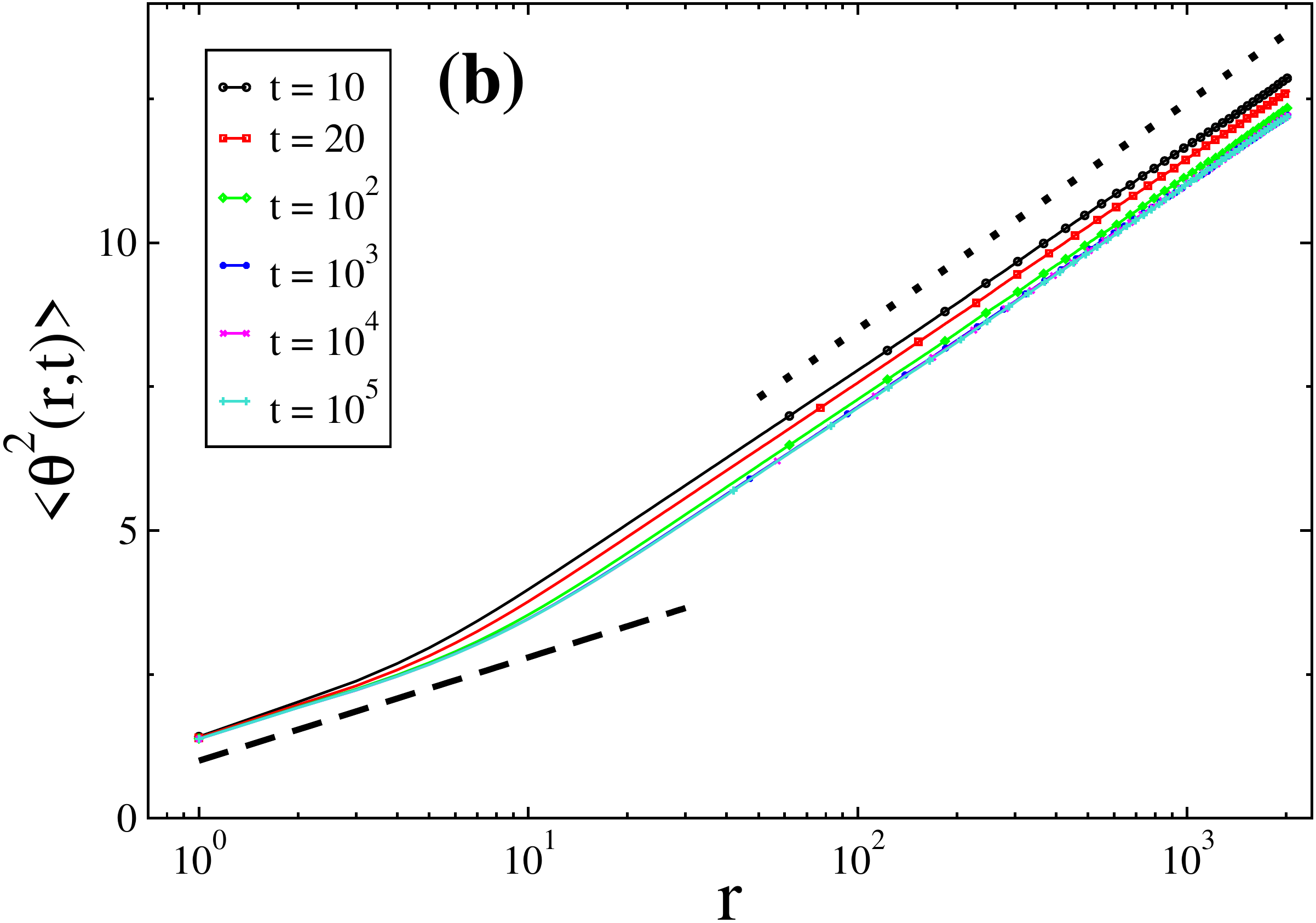}}}
	\caption{The winding angle variance (WAV) vs curvilinear distance $r$ for quenches from 
	infinitely high temperature to (a) $T_c = 0.5$ at $p \simeq 0.107$, and (b) $T_c = 0$ at $p \simeq 0.103$. The system size for datasets at different times (see the key) is $L = 1024$ unless mentioned explicitly. The dashed and dotted lines in both panels represent the stochastic Loewner evolution (SLE) with $\kappa \simeq 1.93$ and $\kappa = 6$, respectively.}
	\label{fig7}
\end{figure}

In Fig.~\ref{fig7}(a), the WAV $\langle \theta^2(r, t) \rangle$ is plotted for a quench from high $T$ to $T_c \simeq 0.5$ at $p \simeq 0.107$. The slope of different curves at small $r$ is consistent with $\kappa \simeq 1.93$. Since the growth of critical correlations is extremely slow, the WAV curve at the longest simulation time ($t=10^5$) for $L = 1024$ agrees with the slope of $\kappa \simeq 1.93$ till $r \simeq 5$ only (at time $t=10^8$ on $L=128$ the slope with $\kappa \simeq 1.93$ remains till $r \simeq 20$). Notice that the value of interfacial fractal dimension $d_{\rm l}$ obtained from the relation~\eqref{frac_dim} is close ($\simeq 1.24$) to that at NP. However, the fractal dimension associated with the cluster area ($d_{\rm A}$) differs significantly. The slope of the WAV curves at large $r$ is consistent with $\kappa \simeq 6$.

In Fig.~\ref{fig7}(b), the behavior of the WAV is explored for a quench from infinitely high $T$ to $T_c = 0$ at $p = p_0$ (strong disorder fixed point). As observed in the previous Section, due to the absence of  thermal fluctuations the dynamics in the system gets ceased soon after the quench. Therefore, we cannot precisely determine the value of $\kappa$ or $d_{\rm l}$ at small growing length scales. However, the initial increase of the WAV in frozen states is enough to point out that the structures/interfaces at $T_c(p_0)$ are not smooth. The interesting fact is that the slope of all curves (starting from as early as $t \simeq 10$) at large $r$ is still consistent with $\kappa = 6$.

The values of the fractal dimension $d_{\rm l}$ and $\kappa$ for different critical quenches investigated in this Section are summarized in Tab.~\ref{tab1}.

\begin{figure}[h!]
\vspace{0.5cm}
	\rotatebox{0}{\resizebox{.45\textwidth}{!}{\includegraphics{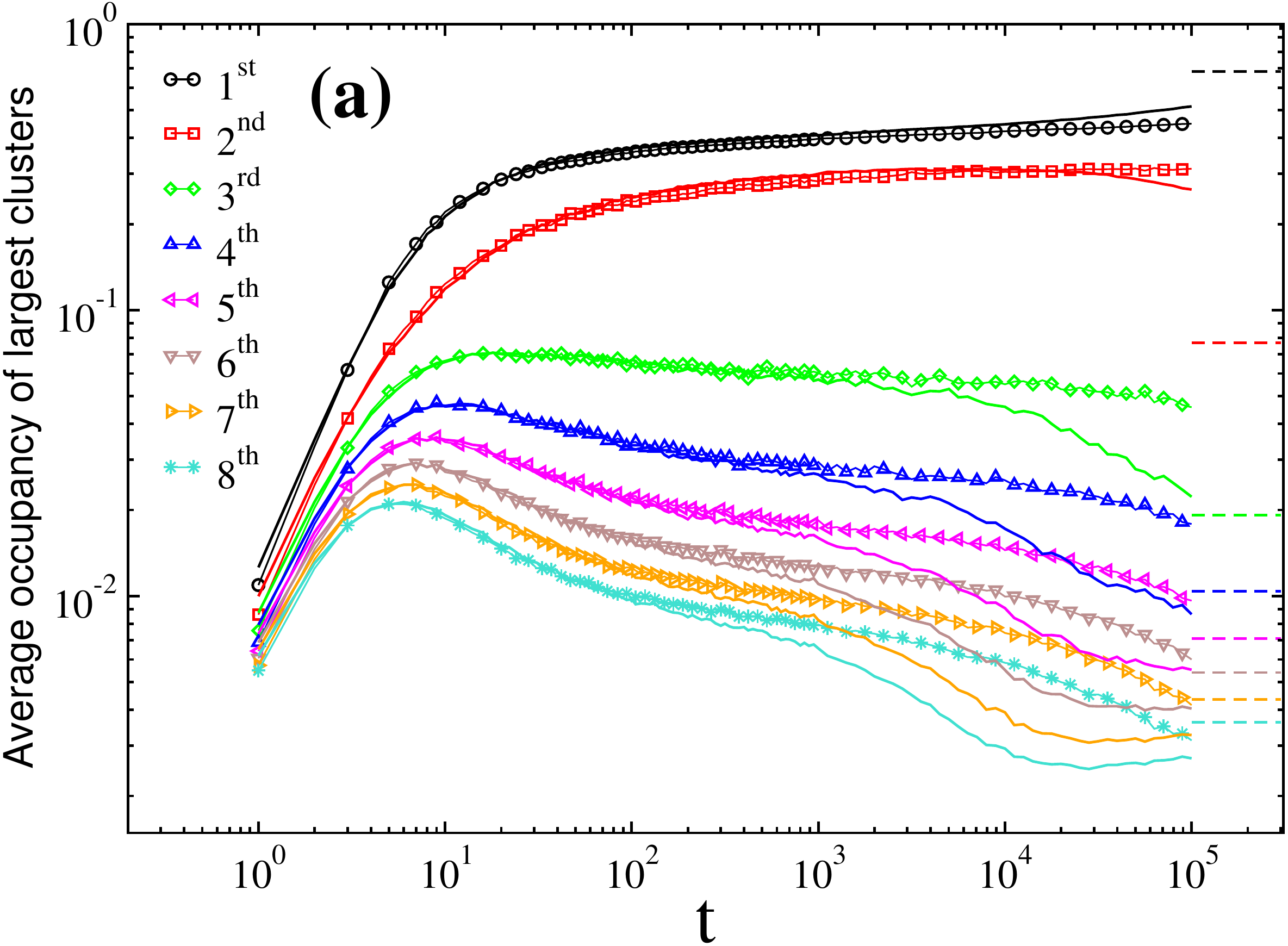}}}
	\rotatebox{0}{\resizebox{.45\textwidth}{!}{\includegraphics{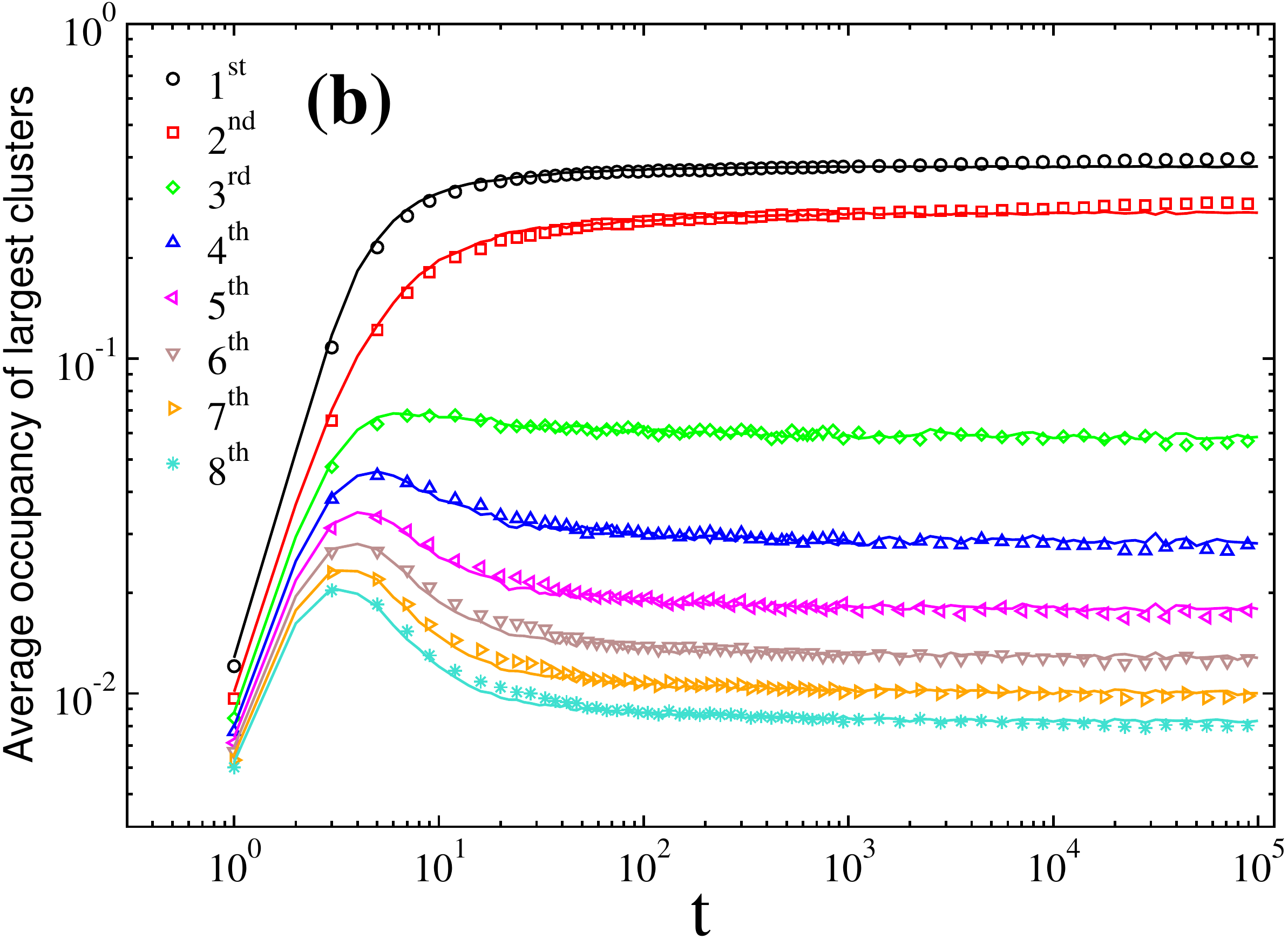}}}
	\caption{The time evolution of the average occupancy rates of the $n^{\rm th}$ largest clusters (see the keys), after a quench from an infinitely high temperature to different critical temperatures. (a) $T_c = 1.687$ (symbols) and $T_{\rm Is}$ (solid lines). The dashed lines indicate the equilibrium values of the average occupancy rates of the $n^{\rm th}$ largest clusters at $T_{\rm Is}$. (b) $T_c = 0.5$ (symbols) and $T_c = 0$ (solid lines).}
	\label{fig8}
\end{figure}

\subsection{Occupancy rates of the largest clusters}

Finally, let us look at the average occupancy rates of the $n^{\rm th}$ largest geometrical clusters (from $1^{\rm st}$ to $8^{\rm th}$) plotted in Fig.~\ref{fig8} for different evolving critical temperatures. Initially, at $t=0$ (high temperature), all clusters are small and are of almost the same mass. With time, large clusters grow at the cost of smaller ones. Asymptotically, the $1^{\rm st}$ largest cluster invades the whole system; however, contrary to a ferromagnetic ground state, the correlation length $\xi_{\rm eq}$ near $T_c$ is infinite. Therefore, other clusters also exist with nonzero probability, which is clearly observed in both panels of the figure. In panel (a) the average occupation rates are presented for $T_c \simeq 1.687$ at $p = 0.07$, and $T_{\rm Is}$ at $p=0$. The initial time evolution of these quantities is similar for both the cases. The deviations from the pure case arise only at late times. In panel (b) the average occupation rates are shown for $T_c \simeq 0.5$ at $p \simeq 0.107$, and $T_c(p_0) = 0$. An important point to learn in this panel is that the time evolution of all these numbers is similar for both quenches, even at the longest timescales.

\section{Summary and discussions}
\label{S5}

The effects of \textit{frustration} on critical phenomena have been a matter of primary concerns in the past few decades. People have been fascinated about how frustration modifies the critical properties. In this context, a special attention is paid to the $2d$ frustrated systems, where a weak disorder often acts as a \textit{marginally} irrelevant perturbation~\cite{doi:10.1142/2460,Dotsenko_1982} to the pure fixed point. Moreover, with increase in the disorder, a rich \textit{multicritical} behavior~\cite{PhysRevB.29.4026,PhysRevLett.61.625} also emerges in these systems. There are numerous studies in this direction; however, most of them are mainly concerned with the static aspects of criticality. The nonequilibrium properties, e.g., dynamical critical exponents, are equally important, and sometimes, they even provide good understanding of the equilibrium properties as well~\cite{RevModPhys.49.435,janssen1989new}.

In this paper we have thoroughly explored the nonequilibrium critical dynamics of the $2d$ $\pm J$ Ising model using large-scale Monte Carlo simulations. Concretely, we followed the evolution of large systems over long periods of time after quenches from different initial conditions to various points on the PF phase boundary, above, below and at the multicritical Nishimori point (NP).

First of all, we investigated the post-quench growth of critical correlations, in terms of the domain growth law, $\xi (t) \sim t^{1/z_c}$, where $\xi (t)$ is the time-dependent correlation length, and $z_c$ is a dynamical critical exponent at the asymptotic timescales. Notice that the $2d$ $\pm J$ Ising model has three fixed points, namely, Ising point at $T = T_{\rm Is}$, NP at $T = T_{\rm N}$, and strong disorder fixed point at $T = T_c(p_0)$. Out of these, the Ising and strong disorder fixed points are attractive, while the NP is of repulsive nature (in the sense of the RG flow). Therefore, it does matter where we are quenching on the PF boundary. Our numerical simulations show that if the quench is made directly to the NP at $T = T_{\rm N}$, soon after the formation of initial critical regions the growth dynamics enters to a long-lasting asymptotic regime, with an asymptotic dynamical exponent $z_c \simeq 6.0$. On the other hand, if the system is quenched above ($T_{\rm N} < T_c(p) < T_{\rm Is}$) or below ($T_c(p_0) < T_c(p) < T_{\rm N}$) the NP, a peculiar scenario is observed --- the dynamics first reaches a preasymptotic regime related to the repulsive fixed point, and later on, it crosses over to the asymptotic regime controlled by the attractive fixed point. However, due to the competition between the different fixed points, a complete crossover remains inaccessible in our simulations.

We also analyzed the dynamical properties of the geometrical features emerged after the critical quenches. For this purpose we mainly exploited the winding angle variance (WAV). This quantity measures a real parameter $\kappa$, which is equivalent to the diffusion parameter of the stochastic Loewner evolution (SLE). We remind that the interfaces in a $2d$ critical system can be envisioned as the random planar curves generated by the SLE with parameter $\kappa$. For small scales $r < \xi(t)$, $\kappa$ attains three distinct values depending on whether $T_c > T_{\rm N}$, $T_c = T_{\rm N}$, or $T_c < T_{\rm N}$, irrespective of the initial quenched state. This exhibits the uniqueness of each universality class on the PF critical boundary. Furthermore, for large scales $r > \xi(t)$, the value of $\kappa$ for all critical quenches from a high temperature phase is consistent with that at the critical random percolation ($\kappa = 6$). Such a behavior onsets beyond an early time $t \sim 10$ and holds till the equilibration time $t_{\rm eq} \sim L^{z_c}$, confirming an emergent critical percolation topology akin to the pure case ($p=0$)~\cite{BlaCorCugPic14,AreBrayCugSic07,Blanchard_2017}.

Before ending, let us discuss some open points and possible future directions. We have seen above that in weakly disordered systems, the access to an asymptotic regime during the critical dynamics turns out to be a challenging problem, at least, from the numerical point of view. Therefore, the analytical efforts are highly encouraged in this direction. Possibly, methods like high-temperature series expansion, which have been applied to the spin glasses too, can be useful tools for the current frustrated system. It is also desirable to extend the present study and use different lattice geometries to verify universality in this respect. Another interesting problem would be to analyze the geometric features at the multicritical point $T_c(p_0) = 0$. Since the single flip Monte Carlo method is non-ergodic at $T = 0$, the exact matching algorithms or the simulated annealing techniques would be of some interests. Finally, we hope that our work will gain some attention among the scientific community and attract other researchers towards these persisting issues in a simplest frustrated system.

\appendix

\section{Short time critical dynamics: crossover in dynamical exponent}
\label{A1}

The short time critical dynamics (STCD) is a unique approach to investigate the universal features of the critical phenomenon. We use it here to calculate the dynamical exponent $z_c$ for a quench from an initially ordered state to different critical temperatures $T_c(p)$ on the paramagnetic-ferromagnetic (PF) line above the Nishimori point (see the main text for the details). We emphasize that for the sensitive determination of the critical exponents using STCD, a quench  from an ordered state is more suitable, as it has less statistical fluctuations.

\begin{figure}[h!]
	\vspace{0.5cm}
	\centering
	\rotatebox{0}{\resizebox{.45\textwidth}{!}{\includegraphics{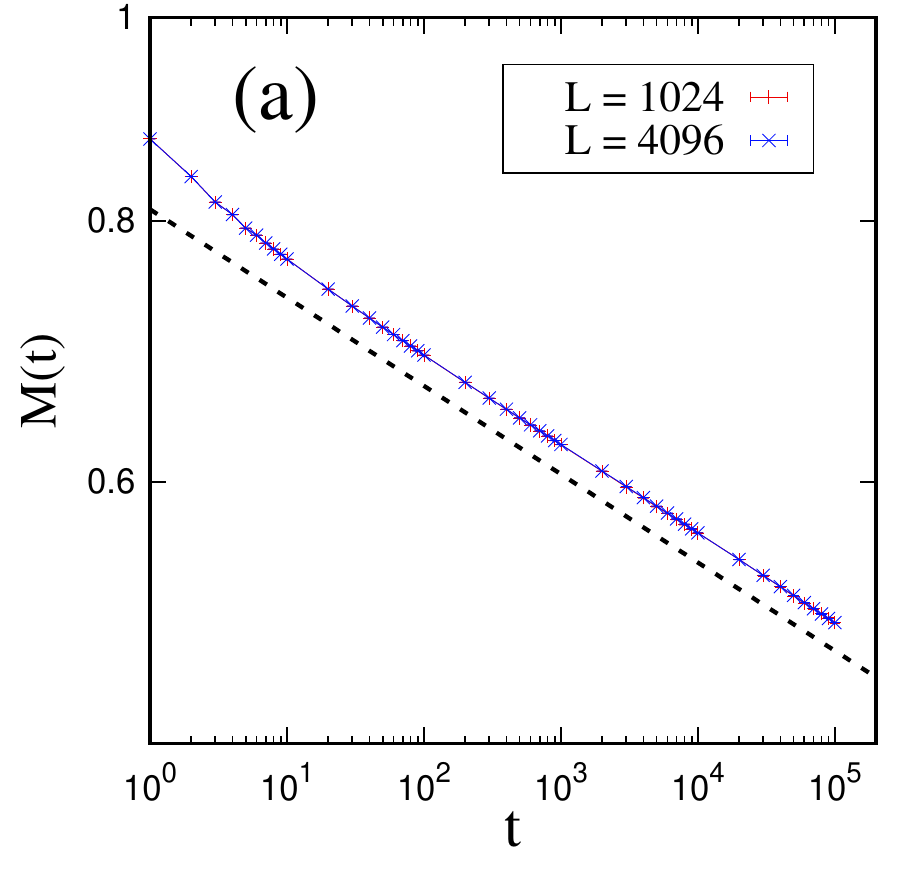}}}
	\rotatebox{0}{\resizebox{.47\textwidth}{!}{\includegraphics{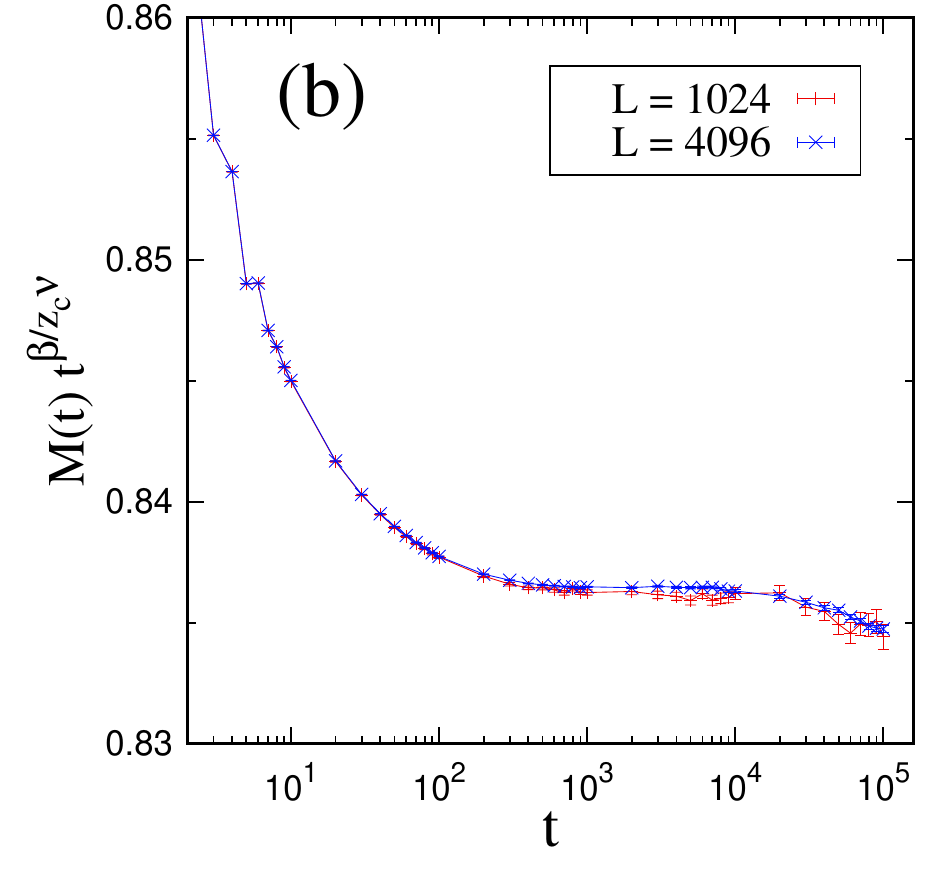}}}
	\caption{(a) $M(t)$ vs $t$ in log-log scale and (b) $M(t) t^{\beta/z_c \nu}$ vs $t$ also in log-linear scale, 
	after a quench from a completely ordered state to $T_c = 1.687$ on the critical line above the Nishimori point,
	 for different system sizes (see the keys). Here, $\beta$ and $\nu$ are the critical exponents of the magnetization and the correlation length, respectively, with the values of the Ising fixed point ($\beta / \nu = 0.125$), and $z_c$ is a preasymptotic value of the dynamical critical exponent ($z_c \simeq 2.96$). The dashed line in (a) denotes the decay law $M(t) \sim t^{-\beta/z_c \nu}$ (see the main text).}
	\label{fig9}
\end{figure}

We take the square lattice system with the linear sizes $L = 1024$ and $L = 4096$, initially prepared in an ordered state by choosing all the spins $+1$. In different simulations the system is evolved at $T_c = 1.875$ with $p=0.05$, $T_c = 1.687$ with $p=0.07$, and $T_c = 1.580$ with $p=0.08$, using the Metropolis algorithm~\eqref{metrop}. To achieve good statistical accuracy  we average the observables over $10000$ independent thermal histories and disordered configurations. The simulations are fastened by implementing an optimized code on the graphics processing unit (GPU).

\begin{figure}[t!]
	\vspace{0.5cm}
	\centering
	\rotatebox{0}{\resizebox{0.75\textwidth}{!}{\includegraphics{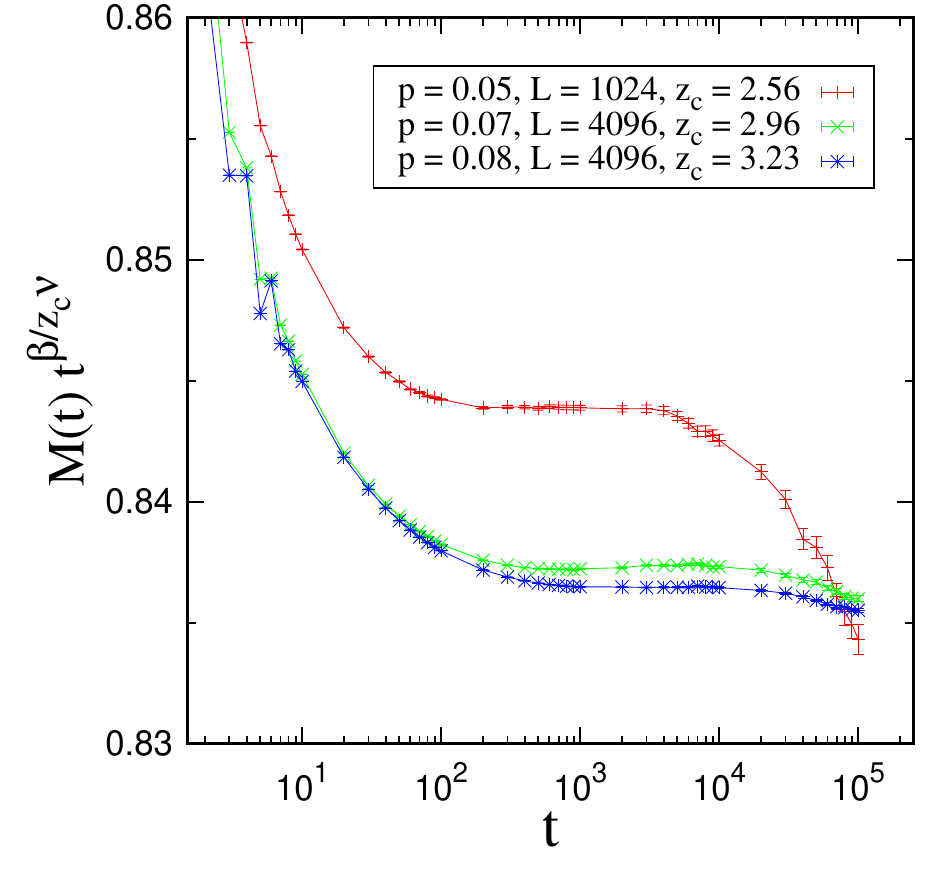}}}
	\caption{Similar to Figure~\ref{fig9}(b), the plot of $M(t) t^{\beta/z_c \nu}$ vs $t$ in log-linear scale, after a quench from a completely ordered state to different critical quenches $T_c(p)$ on the PF line (see the keys)}
	\label{fig10}
\end{figure}

Let us start by presenting the simulation results first for disorder value $p=0.07$. In Fig.~\ref{fig9}(a), the magnetization density $M(t)$ is plotted against time $t$ measured in Monte Carlo step units. After a transient $t \sim 500$, the numerical data for both system sizes are consistent with a power law decay~\eqref{m1}, $M(t) \sim t^{-\beta/z_c \nu}$. A fit in the time-window $t \in [500, 10000]$ gives $z_c \simeq 2.96$, where the ratio $\beta/\nu$ is fixed to the critical Ising value $\beta/\nu= 0.125$. The estimated value of $z_c$ is compatible with the one obtained from the growth of the correlation length (see Tab.~\ref{tab1} in the main text). We further observe in the same figure that at $t > 10^4$, the slope of the magnetization density has slightly increased. To visualize this clearly, we plot in Fig.~\ref{fig9}(b) the re-scaled magnetization density $M(t) t^{\beta/z_c \nu}$ against $t$, where $\beta/z_c \nu \simeq 0.0421$, the value obtained from fit above. In such a kind of plot the plateau for $t \in [500, 10000]$ indicates that the previous exponent $z_c \simeq 2.96$ is the correct one in this regime. However, going beyond $t \sim 10^4$ to the longest simulation time $t = 10^5$, $M(t) t^{\beta/z_c \nu}$ continuously decreases, which indicates that the value of the exponent $z_c$ starts to decrease.

In the longest time scales that we access the dynamics is still in a \textit{crossover}. The value $z_c \simeq 2.96$ obtained from the early time data is preasymptotic. However, to observe the true asymptotic value $z_c \simeq 2.17$, one would need to reach times which go way beyond the ones accessible with these simulations. Similar result is obtained for other disorder values chosen on the PF line above the Nishimori point. In Fig.~\ref{fig10}, we summarize our data for different disorder values on the PF line. As expected, for small disorder value $p=0.05$, a slightly reduced value of preasymptotic exponent ($z_c \simeq 2.56$) is recovered, which starts to decrease on a comparably smaller crossover time ($t\simeq 3000$). On the other hand, for a larger disorder value $p=0.08$, an exponent $z_c \simeq 3.23$ is achieved. The latter also decreases but on the timescale beyond $t > 10^4$.

In the end, we mention that it would be desirable to access large timescales in the current GPU simulations. However, this requires a huge computational effort, which is not feasible with our current resources. For an example, to access timescales up to $t = 10^5$ MCS with $10000$ samples, our simulations took approximately $220$ GPU hours on an NVIDIA GeForce RTX $3080$ graphic card (with $8704$ CUDA cores). To reach timescales up to $t = 10^6$ MCS, a $10$ times larger computational effort would be required.

\section{Dynamical length scale from time-dependent Binder cumulant}
\label{A2}

For a critical quench from the completely ordered state, the dynamical length scale can also be extracted from the time-dependent Binder cumulant $U(t)$ [see Eqs.~\eqref{bind}-\eqref{bind1}]. In the scaling regime, independent of the initial start, the quantity $\left[U(t)\right]^{1/2}$ should be proportional to the correlation length $\xi (t)$ extracted from the decay of the spatial correlation function in the main text.

\begin{figure}[t!]
	\vspace{0.5cm}
	\centering
	\rotatebox{0}{\resizebox{0.85\textwidth}{!}{\includegraphics{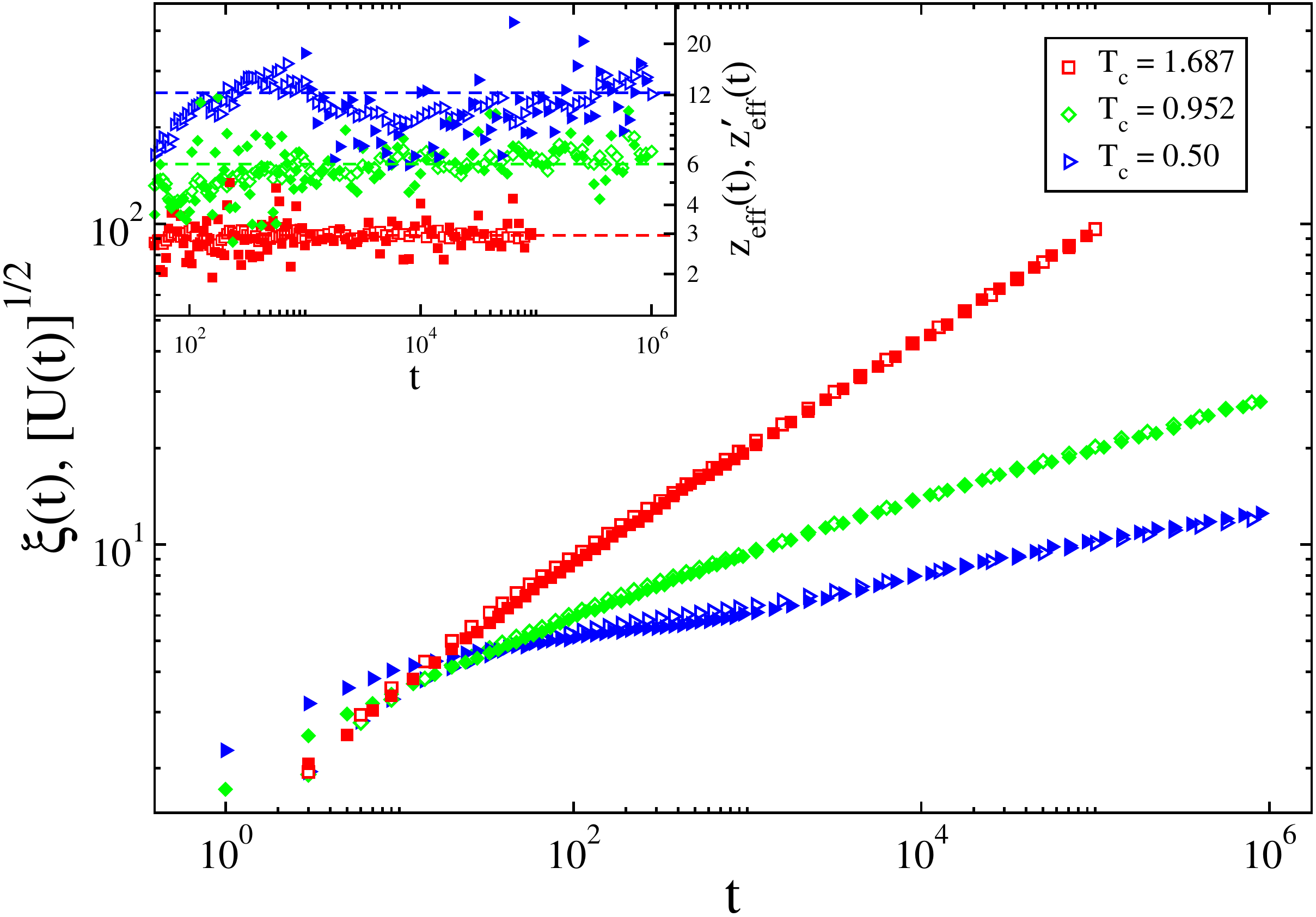}}}
	\caption{Plot of the correlation length $\xi (t)$ and quantity $\left[U(t)\right]^{1/2}$ against time $t$, in log-log scale, for different critical quenches (see the key) from an infinitely high temperature $T \gg T_c$ and a completely ordered state, respectively. The empty symbols denote data for $\xi (t)$, and the filled symbols correspond to $\left[U(t)\right]^{1/2}$. The inset plots the effective dynamical exponents $z_{\rm eff}(t)$ and $z^{'}_{\rm eff}(t)$ against $t$ in log-log scale for the datasets in the main frame (see text for details). The dashed horizontal lines represent the late-time plateau of $z_{\rm eff}$.}
	\label{fig11}
\end{figure}

In Fig.~\ref{fig11}, we compare the quantities $\xi (t)$ and $\left[U(t)\right]^{1/2}$ for different critical quenches from an infinitely high temperature and a completely ordered state, respectively. For critical quench above the Nishimori point $T_{\rm N} = 0.952$, the linear size of the system is $L = 1024$, while for quenches at/below $T_{\rm N}$, the linear size of the system is $L = 128$. The data shown in Fig.~\ref{fig11} is averaged over $5000 - 10000$ independent thermal histories and disordered configurations.

In Fig.~\ref{fig11}, the datasets for $\left[U(t)\right]^{1/2}$ are scaled by appropriate multiplicative prefactors to fall on the datasets for $\xi (t)$. We observe that once the scaling regime is set, $\xi (t)$ and $\left[U(t)\right]^{1/2}$ are in excellent agreement with each other. For rigorousness, we calculate the effective exponent $z^{'}_{\rm eff}(t)$ for $\left[U(t)\right]^{1/2}$ defined as,
\be
\label{z_eff1}
\frac{1}{z^{'}_{\rm eff}(t)} = \frac{d\ln \left[U(t)\right]^{1/2}}{d\ln t}
\; ,
\ee
and we compare it with the exponent $z_{\rm eff}(t)$ [see Eq.~\eqref{z_eff} in main text] of $\xi(t)$ in the inset. The long time trend of $z^{'}_{\rm eff}(t)$ agrees with $z_{\rm eff}(t)$. Therefore, the corresponding estimates of the dynamical critical exponents $z_c$ are also similar (indicated by dashed horizontal lines in inset). We notice that the fluctuations in $z^{'}_{\rm eff}$ are comparatively larger. This is likely because the Binder cumulant $U(t)$ is a macroscopic observable and requires a huge statistics.

\vspace{0.5cm}

\noindent
{\bf Acknowledgements} The authors acknowledge financial support from ANR-19-CE30-0014.

\bibliography{ref}

\begin{thebibliography}{72}%
\makeatletter
\providecommand \@ifxundefined [1]{%
 \@ifx{#1\undefined}
}%
\providecommand \@ifnum [1]{%
 \ifnum #1\expandafter \@firstoftwo
 \else \expandafter \@secondoftwo
 \fi
}%
\providecommand \@ifx [1]{%
 \ifx #1\expandafter \@firstoftwo
 \else \expandafter \@secondoftwo
 \fi
}%
\providecommand \natexlab [1]{#1}%
\providecommand \enquote  [1]{``#1''}%
\providecommand \bibnamefont  [1]{#1}%
\providecommand \bibfnamefont [1]{#1}%
\providecommand \citenamefont [1]{#1}%
\providecommand \href@noop [0]{\@secondoftwo}%
\providecommand \href [0]{\begingroup \@sanitize@url \@href}%
\providecommand \@href[1]{\@@startlink{#1}\@@href}%
\providecommand \@@href[1]{\endgroup#1\@@endlink}%
\providecommand \@sanitize@url [0]{\catcode `\\12\catcode `\$12\catcode
  `\&12\catcode `\#12\catcode `\^12\catcode `\_12\catcode `\%12\relax}%
\providecommand \@@startlink[1]{}%
\providecommand \@@endlink[0]{}%
\providecommand \url  [0]{\begingroup\@sanitize@url \@url }%
\providecommand \@url [1]{\endgroup\@href {#1}{\urlprefix }}%
\providecommand \urlprefix  [0]{URL }%
\providecommand \Eprint [0]{\href }%
\providecommand \doibase [0]{http://dx.doi.org/}%
\providecommand \selectlanguage [0]{\@gobble}%
\providecommand \bibinfo  [0]{\@secondoftwo}%
\providecommand \bibfield  [0]{\@secondoftwo}%
\providecommand \translation [1]{[#1]}%
\providecommand \BibitemOpen [0]{}%
\providecommand \bibitemStop [0]{}%
\providecommand \bibitemNoStop [0]{.\EOS\space}%
\providecommand \EOS [0]{\spacefactor3000\relax}%
\providecommand \BibitemShut  [1]{\csname bibitem#1\endcsname}%
\let\auto@bib@innerbib\@empty
%</preamble>
\bibitem [{\citenamefont {Edwards}\ and\ \citenamefont
  {Anderson}(1975)}]{Edwards_1975}%
  \BibitemOpen
  \bibfield  {author} {\bibinfo {author} {\bibfnamefont {S.~F.}\ \bibnamefont
  {Edwards}}\ and\ \bibinfo {author} {\bibfnamefont {P.~W.}\ \bibnamefont
  {Anderson}},\ }\href {\doibase 10.1088/0305-4608/5/5/017} {\bibfield
  {journal} {\bibinfo  {journal} {J. Phys. F: Metal Phys.}\ }\textbf {\bibinfo
  {volume} {5}},\ \bibinfo {pages} {965} (\bibinfo {year} {1975})}\BibitemShut
  {NoStop}%
\bibitem [{\citenamefont {Sourlas}(1989)}]{sourlas1989spin}%
  \BibitemOpen
  \bibfield  {author} {\bibinfo {author} {\bibfnamefont {N.}~\bibnamefont
  {Sourlas}},\ }\href@noop {} {\bibfield  {journal} {\bibinfo  {journal}
  {Nature}\ }\textbf {\bibinfo {volume} {339}},\ \bibinfo {pages} {693}
  (\bibinfo {year} {1989})}\BibitemShut {NoStop}%
\bibitem [{\citenamefont {Dotsenko}(1995)}]{doi:10.1142/2460}%
  \BibitemOpen
  \bibfield  {author} {\bibinfo {author} {\bibfnamefont {V.}~\bibnamefont
  {Dotsenko}},\ }\href {\doibase 10.1142/2460} {\emph {\bibinfo {title} {An
  Introduction to the Theory of Spin Glasses and Neural Networks}}}\ (\bibinfo
  {publisher} {World Scientific},\ \bibinfo {address} {Singapore},\ \bibinfo
  {year} {1995})\BibitemShut {NoStop}%
\bibitem [{\citenamefont {Nishimori}(2001)}]{nishimori2001statistical}%
  \BibitemOpen
  \bibfield  {author} {\bibinfo {author} {\bibfnamefont {H.}~\bibnamefont
  {Nishimori}},\ }\href@noop {} {\emph {\bibinfo {title} {Statistical physics
  of spin glasses and information processing: an introduction}}},\ \bibinfo
  {number} {111}\ (\bibinfo  {publisher} {Clarendon Press},\ \bibinfo {address}
  {Oxford},\ \bibinfo {year} {2001})\BibitemShut {NoStop}%
\bibitem [{\citenamefont {Kitaev}(1997)}]{Kitaev_1997}%
  \BibitemOpen
  \bibfield  {author} {\bibinfo {author} {\bibfnamefont {A.~Y.}\ \bibnamefont
  {Kitaev}},\ }\href {\doibase 10.1070/RM1997v052n06ABEH002155} {\bibfield
  {journal} {\bibinfo  {journal} {Russian Mathematical Surveys}\ }\textbf
  {\bibinfo {volume} {52}},\ \bibinfo {pages} {1191} (\bibinfo {year}
  {1997})}\BibitemShut {NoStop}%
\bibitem [{\citenamefont {Kitaev}(2003)}]{KITAEV20032}%
  \BibitemOpen
  \bibfield  {author} {\bibinfo {author} {\bibfnamefont {A.}~\bibnamefont
  {Kitaev}},\ }\href {\doibase https://doi.org/10.1016/S0003-4916(02)00018-0}
  {\bibfield  {journal} {\bibinfo  {journal} {Annals of Physics}\ }\textbf
  {\bibinfo {volume} {303}},\ \bibinfo {pages} {2} (\bibinfo {year}
  {2003})}\BibitemShut {NoStop}%
\bibitem [{\citenamefont {Dennis}\ \emph {et~al.}(2002)\citenamefont {Dennis},
  \citenamefont {Kitaev}, \citenamefont {Landahl},\ and\ \citenamefont
  {Preskill}}]{doi:10.1063/1.1499754}%
  \BibitemOpen
  \bibfield  {author} {\bibinfo {author} {\bibfnamefont {E.}~\bibnamefont
  {Dennis}}, \bibinfo {author} {\bibfnamefont {A.}~\bibnamefont {Kitaev}},
  \bibinfo {author} {\bibfnamefont {A.}~\bibnamefont {Landahl}}, \ and\
  \bibinfo {author} {\bibfnamefont {J.}~\bibnamefont {Preskill}},\ }\href
  {\doibase 10.1063/1.1499754} {\bibfield  {journal} {\bibinfo  {journal} {J.
  Math. Phys.}\ }\textbf {\bibinfo {volume} {43}},\ \bibinfo {pages} {4452}
  (\bibinfo {year} {2002})}\BibitemShut {NoStop}%
\bibitem [{\citenamefont {Berthier}\ \emph {et~al.}(2011)\citenamefont
  {Berthier}, \citenamefont {Biroli}, \citenamefont {Bouchaud}, \citenamefont
  {Cipelletti},\ and\ \citenamefont {van Saarloos}}]{berthier2011dynamical}%
  \BibitemOpen
  \bibfield  {author} {\bibinfo {author} {\bibfnamefont {L.}~\bibnamefont
  {Berthier}}, \bibinfo {author} {\bibfnamefont {G.}~\bibnamefont {Biroli}},
  \bibinfo {author} {\bibfnamefont {J.}~\bibnamefont {Bouchaud}}, \bibinfo
  {author} {\bibfnamefont {L.}~\bibnamefont {Cipelletti}}, \ and\ \bibinfo
  {author} {\bibfnamefont {W.}~\bibnamefont {van Saarloos}},\ }\href@noop {}
  {\emph {\bibinfo {title} {Dynamical heterogeneities in glasses, colloids, and
  granular media}}},\ Vol.\ \bibinfo {volume} {150}\ (\bibinfo  {publisher}
  {Oxford University Press},\ \bibinfo {address} {Oxford},\ \bibinfo {year}
  {2011})\BibitemShut {NoStop}%
\bibitem [{\citenamefont {Dotsenko}\ and\ \citenamefont
  {Dotsenko}(1982)}]{Dotsenko_1982}%
  \BibitemOpen
  \bibfield  {author} {\bibinfo {author} {\bibfnamefont {V.~S.}\ \bibnamefont
  {Dotsenko}}\ and\ \bibinfo {author} {\bibfnamefont {V.~S.}\ \bibnamefont
  {Dotsenko}},\ }\href {\doibase 10.1088/0022-3719/15/3/015} {\bibfield
  {journal} {\bibinfo  {journal} {J. Phys. C: Solid State Phys.}\ }\textbf
  {\bibinfo {volume} {15}},\ \bibinfo {pages} {495} (\bibinfo {year}
  {1982})}\BibitemShut {NoStop}%
\bibitem [{\citenamefont {Le~Doussal}\ and\ \citenamefont
  {Harris}(1988)}]{PhysRevLett.61.625}%
  \BibitemOpen
  \bibfield  {author} {\bibinfo {author} {\bibfnamefont {P.}~\bibnamefont
  {Le~Doussal}}\ and\ \bibinfo {author} {\bibfnamefont {A.~B.}\ \bibnamefont
  {Harris}},\ }\href {\doibase 10.1103/PhysRevLett.61.625} {\bibfield
  {journal} {\bibinfo  {journal} {Phys. Rev. Lett.}\ }\textbf {\bibinfo
  {volume} {61}},\ \bibinfo {pages} {625} (\bibinfo {year} {1988})}\BibitemShut
  {NoStop}%
\bibitem [{\citenamefont {Cho}\ and\ \citenamefont
  {Fisher}(1997)}]{PhysRevB.55.1025}%
  \BibitemOpen
  \bibfield  {author} {\bibinfo {author} {\bibfnamefont {S.}~\bibnamefont
  {Cho}}\ and\ \bibinfo {author} {\bibfnamefont {M.~P.~A.}\ \bibnamefont
  {Fisher}},\ }\href {\doibase 10.1103/PhysRevB.55.1025} {\bibfield  {journal}
  {\bibinfo  {journal} {Phys. Rev. B}\ }\textbf {\bibinfo {volume} {55}},\
  \bibinfo {pages} {1025} (\bibinfo {year} {1997})}\BibitemShut {NoStop}%
\bibitem [{\citenamefont {Honecker}\ \emph {et~al.}(2001)\citenamefont
  {Honecker}, \citenamefont {Picco},\ and\ \citenamefont
  {Pujol}}]{PhysRevLett.87.047201}%
  \BibitemOpen
  \bibfield  {author} {\bibinfo {author} {\bibfnamefont {A.}~\bibnamefont
  {Honecker}}, \bibinfo {author} {\bibfnamefont {M.}~\bibnamefont {Picco}}, \
  and\ \bibinfo {author} {\bibfnamefont {P.}~\bibnamefont {Pujol}},\ }\href
  {\doibase 10.1103/PhysRevLett.87.047201} {\bibfield  {journal} {\bibinfo
  {journal} {Phys. Rev. Lett.}\ }\textbf {\bibinfo {volume} {87}},\ \bibinfo
  {pages} {047201} (\bibinfo {year} {2001})}\BibitemShut {NoStop}%
\bibitem [{\citenamefont {Merz}\ and\ \citenamefont
  {Chalker}(2002)}]{PhysRevB.65.054425}%
  \BibitemOpen
  \bibfield  {author} {\bibinfo {author} {\bibfnamefont {F.}~\bibnamefont
  {Merz}}\ and\ \bibinfo {author} {\bibfnamefont {J.~T.}\ \bibnamefont
  {Chalker}},\ }\href {\doibase 10.1103/PhysRevB.65.054425} {\bibfield
  {journal} {\bibinfo  {journal} {Phys. Rev. B}\ }\textbf {\bibinfo {volume}
  {65}},\ \bibinfo {pages} {054425} (\bibinfo {year} {2002})}\BibitemShut
  {NoStop}%
\bibitem [{\citenamefont {Picco}\ \emph {et~al.}(2006)\citenamefont {Picco},
  \citenamefont {Honecker},\ and\ \citenamefont {Pujol}}]{Picco_2006}%
  \BibitemOpen
  \bibfield  {author} {\bibinfo {author} {\bibfnamefont {M.}~\bibnamefont
  {Picco}}, \bibinfo {author} {\bibfnamefont {A.}~\bibnamefont {Honecker}}, \
  and\ \bibinfo {author} {\bibfnamefont {P.}~\bibnamefont {Pujol}},\ }\href
  {\doibase 10.1088/1742-5468/2006/09/P09006} {\bibfield  {journal} {\bibinfo
  {journal} {J. Stat. Mech.}\ }\textbf {\bibinfo {volume} {2006}},\ \bibinfo
  {pages} {P09006} (\bibinfo {year} {2006})}\BibitemShut {NoStop}%
\bibitem [{\citenamefont {de~Queiroz}(2006)}]{PhysRevB.73.064410}%
  \BibitemOpen
  \bibfield  {author} {\bibinfo {author} {\bibfnamefont {S.~L.~A.}\
  \bibnamefont {de~Queiroz}},\ }\href {\doibase 10.1103/PhysRevB.73.064410}
  {\bibfield  {journal} {\bibinfo  {journal} {Phys. Rev. B}\ }\textbf {\bibinfo
  {volume} {73}},\ \bibinfo {pages} {064410} (\bibinfo {year}
  {2006})}\BibitemShut {NoStop}%
\bibitem [{\citenamefont {Ohzeki}(2009)}]{PhysRevE.79.021129}%
  \BibitemOpen
  \bibfield  {author} {\bibinfo {author} {\bibfnamefont {M.}~\bibnamefont
  {Ohzeki}},\ }\href {\doibase 10.1103/PhysRevE.79.021129} {\bibfield
  {journal} {\bibinfo  {journal} {Phys. Rev. E}\ }\textbf {\bibinfo {volume}
  {79}},\ \bibinfo {pages} {021129} (\bibinfo {year} {2009})}\BibitemShut
  {NoStop}%
\bibitem [{\citenamefont {Hasenbusch}\ \emph
  {et~al.}(2008{\natexlab{a}})\citenamefont {Hasenbusch}, \citenamefont
  {Toldin}, \citenamefont {Pelissetto},\ and\ \citenamefont
  {Vicari}}]{PhysRevE.77.051115}%
  \BibitemOpen
  \bibfield  {author} {\bibinfo {author} {\bibfnamefont {M.}~\bibnamefont
  {Hasenbusch}}, \bibinfo {author} {\bibfnamefont {F.~P.}\ \bibnamefont
  {Toldin}}, \bibinfo {author} {\bibfnamefont {A.}~\bibnamefont {Pelissetto}},
  \ and\ \bibinfo {author} {\bibfnamefont {E.}~\bibnamefont {Vicari}},\ }\href
  {\doibase 10.1103/PhysRevE.77.051115} {\bibfield  {journal} {\bibinfo
  {journal} {Phys. Rev. E}\ }\textbf {\bibinfo {volume} {77}},\ \bibinfo
  {pages} {051115} (\bibinfo {year} {2008}{\natexlab{a}})}\BibitemShut
  {NoStop}%
\bibitem [{\citenamefont {Parisen~Toldin}\ \emph {et~al.}(2009)\citenamefont
  {Parisen~Toldin}, \citenamefont {Pelissetto},\ and\ \citenamefont
  {Vicari}}]{parisen2009strong}%
  \BibitemOpen
  \bibfield  {author} {\bibinfo {author} {\bibfnamefont {F.}~\bibnamefont
  {Parisen~Toldin}}, \bibinfo {author} {\bibfnamefont {A.}~\bibnamefont
  {Pelissetto}}, \ and\ \bibinfo {author} {\bibfnamefont {E.}~\bibnamefont
  {Vicari}},\ }\href@noop {} {\bibfield  {journal} {\bibinfo  {journal} {J.
  Stat. Phys.}\ }\textbf {\bibinfo {volume} {135}},\ \bibinfo {pages} {1039}
  (\bibinfo {year} {2009})}\BibitemShut {NoStop}%
\bibitem [{\citenamefont {Hasenbusch}\ \emph
  {et~al.}(2008{\natexlab{b}})\citenamefont {Hasenbusch}, \citenamefont
  {Toldin}, \citenamefont {Pelissetto},\ and\ \citenamefont
  {Vicari}}]{PhysRevE.78.011110}%
  \BibitemOpen
  \bibfield  {author} {\bibinfo {author} {\bibfnamefont {M.}~\bibnamefont
  {Hasenbusch}}, \bibinfo {author} {\bibfnamefont {F.~P.}\ \bibnamefont
  {Toldin}}, \bibinfo {author} {\bibfnamefont {A.}~\bibnamefont {Pelissetto}},
  \ and\ \bibinfo {author} {\bibfnamefont {E.}~\bibnamefont {Vicari}},\ }\href
  {\doibase 10.1103/PhysRevE.78.011110} {\bibfield  {journal} {\bibinfo
  {journal} {Phys. Rev. E}\ }\textbf {\bibinfo {volume} {78}},\ \bibinfo
  {pages} {011110} (\bibinfo {year} {2008}{\natexlab{b}})}\BibitemShut
  {NoStop}%
\bibitem [{\citenamefont {Wang}\ \emph {et~al.}(2003)\citenamefont {Wang},
  \citenamefont {Harrington},\ and\ \citenamefont {Preskill}}]{WANG200331}%
  \BibitemOpen
  \bibfield  {author} {\bibinfo {author} {\bibfnamefont {C.}~\bibnamefont
  {Wang}}, \bibinfo {author} {\bibfnamefont {J.}~\bibnamefont {Harrington}}, \
  and\ \bibinfo {author} {\bibfnamefont {J.}~\bibnamefont {Preskill}},\ }\href
  {\doibase https://doi.org/10.1016/S0003-4916(02)00019-2} {\bibfield
  {journal} {\bibinfo  {journal} {Ann. Phys.}\ }\textbf {\bibinfo {volume}
  {303}},\ \bibinfo {pages} {31} (\bibinfo {year} {2003})}\BibitemShut
  {NoStop}%
\bibitem [{\citenamefont {Katzgraber}\ \emph {et~al.}(2009)\citenamefont
  {Katzgraber}, \citenamefont {Bombin},\ and\ \citenamefont
  {Mart\'{i}n-Delgado}}]{PhysRevLett.103.090501}%
  \BibitemOpen
  \bibfield  {author} {\bibinfo {author} {\bibfnamefont {H.~G.}\ \bibnamefont
  {Katzgraber}}, \bibinfo {author} {\bibfnamefont {H.}~\bibnamefont {Bombin}},
  \ and\ \bibinfo {author} {\bibfnamefont {M.~A.}\ \bibnamefont
  {Mart\'{i}n-Delgado}},\ }\href {\doibase 10.1103/PhysRevLett.103.090501}
  {\bibfield  {journal} {\bibinfo  {journal} {Phys. Rev. Lett.}\ }\textbf
  {\bibinfo {volume} {103}},\ \bibinfo {pages} {090501} (\bibinfo {year}
  {2009})}\BibitemShut {NoStop}%
\bibitem [{\citenamefont {Kubica}\ \emph {et~al.}(2018)\citenamefont {Kubica},
  \citenamefont {Beverland}, \citenamefont {Brandao}, \citenamefont
  {Preskill},\ and\ \citenamefont {Svore}}]{PhysRevLett.120.180501}%
  \BibitemOpen
  \bibfield  {author} {\bibinfo {author} {\bibfnamefont {A.}~\bibnamefont
  {Kubica}}, \bibinfo {author} {\bibfnamefont {M.~E.}\ \bibnamefont
  {Beverland}}, \bibinfo {author} {\bibfnamefont {F.}~\bibnamefont {Brandao}},
  \bibinfo {author} {\bibfnamefont {J.}~\bibnamefont {Preskill}}, \ and\
  \bibinfo {author} {\bibfnamefont {K.~M.}\ \bibnamefont {Svore}},\ }\href
  {\doibase 10.1103/PhysRevLett.120.180501} {\bibfield  {journal} {\bibinfo
  {journal} {Phys. Rev. Lett.}\ }\textbf {\bibinfo {volume} {120}},\ \bibinfo
  {pages} {180501} (\bibinfo {year} {2018})}\BibitemShut {NoStop}%
\bibitem [{\citenamefont {Nishimori}(1981)}]{10.1143/PTP.66.1169}%
  \BibitemOpen
  \bibfield  {author} {\bibinfo {author} {\bibfnamefont {H.}~\bibnamefont
  {Nishimori}},\ }\href {\doibase 10.1143/PTP.66.1169} {\bibfield  {journal}
  {\bibinfo  {journal} {Prog. Theor. Phys.}\ }\textbf {\bibinfo {volume}
  {66}},\ \bibinfo {pages} {1169} (\bibinfo {year} {1981})}\BibitemShut
  {NoStop}%
\bibitem [{\citenamefont {McMillan}(1984)}]{PhysRevB.29.4026}%
  \BibitemOpen
  \bibfield  {author} {\bibinfo {author} {\bibfnamefont {W.~L.}\ \bibnamefont
  {McMillan}},\ }\href {\doibase 10.1103/PhysRevB.29.4026} {\bibfield
  {journal} {\bibinfo  {journal} {Phys. Rev. B}\ }\textbf {\bibinfo {volume}
  {29}},\ \bibinfo {pages} {4026} (\bibinfo {year} {1984})}\BibitemShut
  {NoStop}%
\bibitem [{\citenamefont {Amoruso}\ and\ \citenamefont
  {Hartmann}(2004)}]{PhysRevB.70.134425}%
  \BibitemOpen
  \bibfield  {author} {\bibinfo {author} {\bibfnamefont {C.}~\bibnamefont
  {Amoruso}}\ and\ \bibinfo {author} {\bibfnamefont {A.~K.}\ \bibnamefont
  {Hartmann}},\ }\href {\doibase 10.1103/PhysRevB.70.134425} {\bibfield
  {journal} {\bibinfo  {journal} {Phys. Rev. B}\ }\textbf {\bibinfo {volume}
  {70}},\ \bibinfo {pages} {134425} (\bibinfo {year} {2004})}\BibitemShut
  {NoStop}%
\bibitem [{\citenamefont {Singh}\ and\ \citenamefont
  {Adler}(1996)}]{PhysRevB.54.364}%
  \BibitemOpen
  \bibfield  {author} {\bibinfo {author} {\bibfnamefont {R.~R.~P.}\
  \bibnamefont {Singh}}\ and\ \bibinfo {author} {\bibfnamefont
  {J.}~\bibnamefont {Adler}},\ }\href {\doibase 10.1103/PhysRevB.54.364}
  {\bibfield  {journal} {\bibinfo  {journal} {Phys. Rev. B}\ }\textbf {\bibinfo
  {volume} {54}},\ \bibinfo {pages} {364} (\bibinfo {year} {1996})}\BibitemShut
  {NoStop}%
\bibitem [{\citenamefont {Aarao~Reis}\ \emph {et~al.}(1999)\citenamefont
  {Aarao~Reis}, \citenamefont {de~Queiroz},\ and\ \citenamefont {dos
  Santos}}]{PhysRevB.60.6740}%
  \BibitemOpen
  \bibfield  {author} {\bibinfo {author} {\bibfnamefont {F.~D.~A.}\
  \bibnamefont {Aarao~Reis}}, \bibinfo {author} {\bibfnamefont {S.~L.~A.}\
  \bibnamefont {de~Queiroz}}, \ and\ \bibinfo {author} {\bibfnamefont {R.~R.}\
  \bibnamefont {dos Santos}},\ }\href {\doibase 10.1103/PhysRevB.60.6740}
  {\bibfield  {journal} {\bibinfo  {journal} {Phys. Rev. B}\ }\textbf {\bibinfo
  {volume} {60}},\ \bibinfo {pages} {6740} (\bibinfo {year}
  {1999})}\BibitemShut {NoStop}%
\bibitem [{\citenamefont {Hohenberg}\ and\ \citenamefont
  {Halperin}(1977)}]{RevModPhys.49.435}%
  \BibitemOpen
  \bibfield  {author} {\bibinfo {author} {\bibfnamefont {P.~C.}\ \bibnamefont
  {Hohenberg}}\ and\ \bibinfo {author} {\bibfnamefont {B.~I.}\ \bibnamefont
  {Halperin}},\ }\href {\doibase 10.1103/RevModPhys.49.435} {\bibfield
  {journal} {\bibinfo  {journal} {Rev. Mod. Phys.}\ }\textbf {\bibinfo {volume}
  {49}},\ \bibinfo {pages} {435} (\bibinfo {year} {1977})}\BibitemShut
  {NoStop}%
\bibitem [{\citenamefont {Tauber}(2014)}]{tauber_2014}%
  \BibitemOpen
  \bibfield  {author} {\bibinfo {author} {\bibfnamefont {U.~C.}\ \bibnamefont
  {Tauber}},\ }\href {\doibase 10.1017/CBO9781139046213} {\emph {\bibinfo
  {title} {Critical Dynamics: A Field Theory Approach to Equilibrium and
  Non-Equilibrium Scaling Behavior}}}\ (\bibinfo  {publisher} {Cambridge
  University Press},\ \bibinfo {address} {Cambridge},\ \bibinfo {year}
  {2014})\BibitemShut {NoStop}%
\bibitem [{\citenamefont {Janssen}\ \emph {et~al.}(1989)\citenamefont
  {Janssen}, \citenamefont {Schaub},\ and\ \citenamefont
  {Schmittmann}}]{janssen1989new}%
  \BibitemOpen
  \bibfield  {author} {\bibinfo {author} {\bibfnamefont {H.}~\bibnamefont
  {Janssen}}, \bibinfo {author} {\bibfnamefont {B.}~\bibnamefont {Schaub}}, \
  and\ \bibinfo {author} {\bibfnamefont {B.}~\bibnamefont {Schmittmann}},\
  }\href@noop {} {\bibfield  {journal} {\bibinfo  {journal} {Z. Phys. B: Cond.
  Matt.}\ }\textbf {\bibinfo {volume} {73}},\ \bibinfo {pages} {539} (\bibinfo
  {year} {1989})}\BibitemShut {NoStop}%
\bibitem [{\citenamefont {Huse}(1989)}]{PhysRevB.40.304}%
  \BibitemOpen
  \bibfield  {author} {\bibinfo {author} {\bibfnamefont {D.~A.}\ \bibnamefont
  {Huse}},\ }\href {\doibase 10.1103/PhysRevB.40.304} {\bibfield  {journal}
  {\bibinfo  {journal} {Phys. Rev. B}\ }\textbf {\bibinfo {volume} {40}},\
  \bibinfo {pages} {304} (\bibinfo {year} {1989})}\BibitemShut {NoStop}%
\bibitem [{\citenamefont {Majumdar}\ \emph {et~al.}(1996)\citenamefont
  {Majumdar}, \citenamefont {Bray}, \citenamefont {Cornell},\ and\
  \citenamefont {Sire}}]{PhysRevLett.77.3704}%
  \BibitemOpen
  \bibfield  {author} {\bibinfo {author} {\bibfnamefont {S.~N.}\ \bibnamefont
  {Majumdar}}, \bibinfo {author} {\bibfnamefont {A.~J.}\ \bibnamefont {Bray}},
  \bibinfo {author} {\bibfnamefont {S.~J.}\ \bibnamefont {Cornell}}, \ and\
  \bibinfo {author} {\bibfnamefont {C.}~\bibnamefont {Sire}},\ }\href {\doibase
  10.1103/PhysRevLett.77.3704} {\bibfield  {journal} {\bibinfo  {journal}
  {Phys. Rev. Lett.}\ }\textbf {\bibinfo {volume} {77}},\ \bibinfo {pages}
  {3704} (\bibinfo {year} {1996})}\BibitemShut {NoStop}%
\bibitem [{\citenamefont {Zheng}(1998)}]{doi:10.1142/S021797929800288X}%
  \BibitemOpen
  \bibfield  {author} {\bibinfo {author} {\bibfnamefont {B.}~\bibnamefont
  {Zheng}},\ }\href {\doibase 10.1142/S021797929800288X} {\bibfield  {journal}
  {\bibinfo  {journal} {Int. J. Mod. Phys. B}\ }\textbf {\bibinfo {volume}
  {12}},\ \bibinfo {pages} {1419} (\bibinfo {year} {1998})}\BibitemShut
  {NoStop}%
\bibitem [{\citenamefont {Godreche}\ and\ \citenamefont
  {Luck}(2002)}]{Godreche_2002}%
  \BibitemOpen
  \bibfield  {author} {\bibinfo {author} {\bibfnamefont {C.}~\bibnamefont
  {Godreche}}\ and\ \bibinfo {author} {\bibfnamefont {J.~M.}\ \bibnamefont
  {Luck}},\ }\href {\doibase 10.1088/0953-8984/14/7/316} {\bibfield  {journal}
  {\bibinfo  {journal} {J. Phys.: Condens. Matter}\ }\textbf {\bibinfo {volume}
  {14}},\ \bibinfo {pages} {1589} (\bibinfo {year} {2002})}\BibitemShut
  {NoStop}%
\bibitem [{\citenamefont {Rosov}\ \emph {et~al.}(1992)\citenamefont {Rosov},
  \citenamefont {Hohenemser},\ and\ \citenamefont
  {Eibsch\"{u}tz}}]{PhysRevB.46.3452}%
  \BibitemOpen
  \bibfield  {author} {\bibinfo {author} {\bibfnamefont {N.}~\bibnamefont
  {Rosov}}, \bibinfo {author} {\bibfnamefont {C.}~\bibnamefont {Hohenemser}}, \
  and\ \bibinfo {author} {\bibfnamefont {M.}~\bibnamefont {Eibsch\"{u}tz}},\
  }\href {\doibase 10.1103/PhysRevB.46.3452} {\bibfield  {journal} {\bibinfo
  {journal} {Phys. Rev. B}\ }\textbf {\bibinfo {volume} {46}},\ \bibinfo
  {pages} {3452} (\bibinfo {year} {1992})}\BibitemShut {NoStop}%
\bibitem [{\citenamefont {Barrett}(1986)}]{PhysRevB.34.3513}%
  \BibitemOpen
  \bibfield  {author} {\bibinfo {author} {\bibfnamefont {P.~H.}\ \bibnamefont
  {Barrett}},\ }\href {\doibase 10.1103/PhysRevB.34.3513} {\bibfield  {journal}
  {\bibinfo  {journal} {Phys. Rev. B}\ }\textbf {\bibinfo {volume} {34}},\
  \bibinfo {pages} {3513} (\bibinfo {year} {1986})}\BibitemShut {NoStop}%
\bibitem [{Note1()}]{Note1}%
  \BibitemOpen
  \bibinfo {note} {Apart from the static critical exponents the dynamical
  critical exponents are also equally important to characterize a critical
  phenomenon. Other nontrivial dynamical exponents are $\lambda _c$ (exponent
  associated with two-time correlations) and $\theta _c$ (persistence
  exponent).}\BibitemShut {Stop}%
\bibitem [{\citenamefont {Wang}\ and\ \citenamefont
  {Hu}(1997)}]{PhysRevE.56.2310}%
  \BibitemOpen
  \bibfield  {author} {\bibinfo {author} {\bibfnamefont {F.-G.}\ \bibnamefont
  {Wang}}\ and\ \bibinfo {author} {\bibfnamefont {C.-K.}\ \bibnamefont {Hu}},\
  }\href {\doibase 10.1103/PhysRevE.56.2310} {\bibfield  {journal} {\bibinfo
  {journal} {Phys. Rev. E}\ }\textbf {\bibinfo {volume} {56}},\ \bibinfo
  {pages} {2310} (\bibinfo {year} {1997})}\BibitemShut {NoStop}%
\bibitem [{\citenamefont {Nightingale}\ and\ \citenamefont
  {Bl\"{o}te}(2000)}]{PhysRevB.62.1089}%
  \BibitemOpen
  \bibfield  {author} {\bibinfo {author} {\bibfnamefont {M.~P.}\ \bibnamefont
  {Nightingale}}\ and\ \bibinfo {author} {\bibfnamefont {H.~W.~J.}\
  \bibnamefont {Bl\"{o}te}},\ }\href {\doibase 10.1103/PhysRevB.62.1089}
  {\bibfield  {journal} {\bibinfo  {journal} {Phys. Rev. B}\ }\textbf {\bibinfo
  {volume} {62}},\ \bibinfo {pages} {1089} (\bibinfo {year}
  {2000})}\BibitemShut {NoStop}%
\bibitem [{\citenamefont {Blanchard}\ \emph {et~al.}(2012)\citenamefont
  {Blanchard}, \citenamefont {Cugliandolo},\ and\ \citenamefont
  {Picco}}]{Blanchard_2012}%
  \BibitemOpen
  \bibfield  {author} {\bibinfo {author} {\bibfnamefont {T.}~\bibnamefont
  {Blanchard}}, \bibinfo {author} {\bibfnamefont {L.~F.}\ \bibnamefont
  {Cugliandolo}}, \ and\ \bibinfo {author} {\bibfnamefont {M.}~\bibnamefont
  {Picco}},\ }\href {\doibase 10.1088/1742-5468/2012/05/P05026} {\bibfield
  {journal} {\bibinfo  {journal} {J. Stat. Mech.}\ }\textbf {\bibinfo {volume}
  {2012}},\ \bibinfo {pages} {P05026} (\bibinfo {year} {2012})}\BibitemShut
  {NoStop}%
\bibitem [{\citenamefont {Ricateau}\ \emph {et~al.}(2018)\citenamefont
  {Ricateau}, \citenamefont {Cugliandolo},\ and\ \citenamefont
  {Picco}}]{Ricateau_2018}%
  \BibitemOpen
  \bibfield  {author} {\bibinfo {author} {\bibfnamefont {H.}~\bibnamefont
  {Ricateau}}, \bibinfo {author} {\bibfnamefont {L.~F.}\ \bibnamefont
  {Cugliandolo}}, \ and\ \bibinfo {author} {\bibfnamefont {M.}~\bibnamefont
  {Picco}},\ }\href {\doibase 10.1088/1742-5468/aa9bb4} {\bibfield  {journal}
  {\bibinfo  {journal} {J. Stat. Mech.}\ }\textbf {\bibinfo {volume} {2018}},\
  \bibinfo {pages} {013201} (\bibinfo {year} {2018})}\BibitemShut {NoStop}%
\bibitem [{\citenamefont {Achiam}\ and\ \citenamefont
  {Kosterlitz}(1978)}]{PhysRevLett.41.128}%
  \BibitemOpen
  \bibfield  {author} {\bibinfo {author} {\bibfnamefont {Y.}~\bibnamefont
  {Achiam}}\ and\ \bibinfo {author} {\bibfnamefont {J.~M.}\ \bibnamefont
  {Kosterlitz}},\ }\href {\doibase 10.1103/PhysRevLett.41.128} {\bibfield
  {journal} {\bibinfo  {journal} {Phys. Rev. Lett.}\ }\textbf {\bibinfo
  {volume} {41}},\ \bibinfo {pages} {128} (\bibinfo {year} {1978})}\BibitemShut
  {NoStop}%
\bibitem [{\citenamefont {Mazenko}\ and\ \citenamefont
  {Valls}(1981)}]{PhysRevB.24.1419}%
  \BibitemOpen
  \bibfield  {author} {\bibinfo {author} {\bibfnamefont {G.~F.}\ \bibnamefont
  {Mazenko}}\ and\ \bibinfo {author} {\bibfnamefont {O.~T.}\ \bibnamefont
  {Valls}},\ }\href {\doibase 10.1103/PhysRevB.24.1419} {\bibfield  {journal}
  {\bibinfo  {journal} {Phys. Rev. B}\ }\textbf {\bibinfo {volume} {24}},\
  \bibinfo {pages} {1419} (\bibinfo {year} {1981})}\BibitemShut {NoStop}%
\bibitem [{\citenamefont {Wang}(1993)}]{PhysRevB.47.869}%
  \BibitemOpen
  \bibfield  {author} {\bibinfo {author} {\bibfnamefont {J.}~\bibnamefont
  {Wang}},\ }\href {\doibase 10.1103/PhysRevB.47.869} {\bibfield  {journal}
  {\bibinfo  {journal} {Phys. Rev. B}\ }\textbf {\bibinfo {volume} {47}},\
  \bibinfo {pages} {869} (\bibinfo {year} {1993})}\BibitemShut {NoStop}%
\bibitem [{\citenamefont {Poole}\ and\ \citenamefont {Jan}(1990)}]{Poole_1990}%
  \BibitemOpen
  \bibfield  {author} {\bibinfo {author} {\bibfnamefont {P.~H.}\ \bibnamefont
  {Poole}}\ and\ \bibinfo {author} {\bibfnamefont {N.}~\bibnamefont {Jan}},\
  }\href {\doibase 10.1088/0305-4470/23/9/009} {\bibfield  {journal} {\bibinfo
  {journal} {J. Phys. A: Math. Gen.}\ }\textbf {\bibinfo {volume} {23}},\
  \bibinfo {pages} {L453} (\bibinfo {year} {1990})}\BibitemShut {NoStop}%
\bibitem [{\citenamefont {Ozeki}\ and\ \citenamefont {Ito}(2007)}]{Ozeki_2007}%
  \BibitemOpen
  \bibfield  {author} {\bibinfo {author} {\bibfnamefont {Y.}~\bibnamefont
  {Ozeki}}\ and\ \bibinfo {author} {\bibfnamefont {N.}~\bibnamefont {Ito}},\
  }\href {\doibase 10.1088/1751-8113/40/31/R01} {\bibfield  {journal} {\bibinfo
   {journal} {J. Phys. A: Math. Theor.}\ }\textbf {\bibinfo {volume} {40}},\
  \bibinfo {pages} {R149} (\bibinfo {year} {2007})}\BibitemShut {NoStop}%
\bibitem [{\citenamefont {Ozeki}\ \emph {et~al.}(2012)\citenamefont {Ozeki},
  \citenamefont {Yotsuyanagi},\ and\ \citenamefont
  {Ito}}]{doi:10.1143/JPSJ.81.074602}%
  \BibitemOpen
  \bibfield  {author} {\bibinfo {author} {\bibfnamefont {Y.}~\bibnamefont
  {Ozeki}}, \bibinfo {author} {\bibfnamefont {S.}~\bibnamefont {Yotsuyanagi}},
  \ and\ \bibinfo {author} {\bibfnamefont {N.}~\bibnamefont {Ito}},\ }\href
  {\doibase 10.1143/JPSJ.81.074602} {\bibfield  {journal} {\bibinfo  {journal}
  {J. Phys. Soc. Japan}\ }\textbf {\bibinfo {volume} {81}},\ \bibinfo {pages}
  {074602} (\bibinfo {year} {2012})}\BibitemShut {NoStop}%
\bibitem [{\citenamefont {Newman}\ and\ \citenamefont
  {Barkema}(1999)}]{newman1999monte}%
  \BibitemOpen
  \bibfield  {author} {\bibinfo {author} {\bibfnamefont {M.}~\bibnamefont
  {Newman}}\ and\ \bibinfo {author} {\bibfnamefont {G.}~\bibnamefont
  {Barkema}},\ }\href@noop {} {\emph {\bibinfo {title} {Monte Carlo Methods in
  Statistical Physics}}}\ (\bibinfo  {publisher} {Oxford University Press, New
  York},\ \bibinfo {year} {1999})\BibitemShut {NoStop}%
\bibitem [{\citenamefont {Wolff}(1989)}]{PhysRevLett.62.361}%
  \BibitemOpen
  \bibfield  {author} {\bibinfo {author} {\bibfnamefont {U.}~\bibnamefont
  {Wolff}},\ }\href {\doibase 10.1103/PhysRevLett.62.361} {\bibfield  {journal}
  {\bibinfo  {journal} {Phys. Rev. Lett.}\ }\textbf {\bibinfo {volume} {62}},\
  \bibinfo {pages} {361} (\bibinfo {year} {1989})}\BibitemShut {NoStop}%
\bibitem [{\citenamefont {Blanchard}\ \emph {et~al.}(2014)\citenamefont
  {Blanchard}, \citenamefont {Corberi}, \citenamefont {Cugliandolo},\ and\
  \citenamefont {Picco}}]{BlaCorCugPic14}%
  \BibitemOpen
  \bibfield  {author} {\bibinfo {author} {\bibfnamefont {T.}~\bibnamefont
  {Blanchard}}, \bibinfo {author} {\bibfnamefont {F.}~\bibnamefont {Corberi}},
  \bibinfo {author} {\bibfnamefont {L.~F.}\ \bibnamefont {Cugliandolo}}, \ and\
  \bibinfo {author} {\bibfnamefont {M.}~\bibnamefont {Picco}},\ }\href
  {\doibase 10.1209/0295-5075/106/66001} {\bibfield  {journal} {\bibinfo
  {journal} {EPL}\ }\textbf {\bibinfo {volume} {106}},\ \bibinfo {pages}
  {66001} (\bibinfo {year} {2014})}\BibitemShut {NoStop}%
\bibitem [{\citenamefont {Arenzon}\ \emph {et~al.}(2007)\citenamefont
  {Arenzon}, \citenamefont {Bray}, \citenamefont {Cugliandolo},\ and\
  \citenamefont {Sicilia}}]{AreBrayCugSic07}%
  \BibitemOpen
  \bibfield  {author} {\bibinfo {author} {\bibfnamefont {J.~J.}\ \bibnamefont
  {Arenzon}}, \bibinfo {author} {\bibfnamefont {A.~J.}\ \bibnamefont {Bray}},
  \bibinfo {author} {\bibfnamefont {L.~F.}\ \bibnamefont {Cugliandolo}}, \ and\
  \bibinfo {author} {\bibfnamefont {A.}~\bibnamefont {Sicilia}},\ }\href
  {\doibase 10.1103/PhysRevLett.98.145701} {\bibfield  {journal} {\bibinfo
  {journal} {Phys. Rev. Lett.}\ }\textbf {\bibinfo {volume} {98}},\ \bibinfo
  {pages} {145701} (\bibinfo {year} {2007})}\BibitemShut {NoStop}%
\bibitem [{\citenamefont {Blanchard}\ \emph {et~al.}(2017)\citenamefont
  {Blanchard}, \citenamefont {Cugliandolo}, \citenamefont {Picco},\ and\
  \citenamefont {Tartaglia}}]{Blanchard_2017}%
  \BibitemOpen
  \bibfield  {author} {\bibinfo {author} {\bibfnamefont {T.}~\bibnamefont
  {Blanchard}}, \bibinfo {author} {\bibfnamefont {L.~F.}\ \bibnamefont
  {Cugliandolo}}, \bibinfo {author} {\bibfnamefont {M.}~\bibnamefont {Picco}},
  \ and\ \bibinfo {author} {\bibfnamefont {A.}~\bibnamefont {Tartaglia}},\
  }\href {\doibase 10.1088/1742-5468/aa9348} {\bibfield  {journal} {\bibinfo
  {journal} {J. Stat. Mech.}\ }\textbf {\bibinfo {volume} {2017}},\ \bibinfo
  {pages} {113201} (\bibinfo {year} {2017})}\BibitemShut {NoStop}%
\bibitem [{\citenamefont {Corberi}\ \emph {et~al.}(2017)\citenamefont
  {Corberi}, \citenamefont {Cugliandolo}, \citenamefont {Insalata},\ and\
  \citenamefont {Picco}}]{CorCugInsPic17}%
  \BibitemOpen
  \bibfield  {author} {\bibinfo {author} {\bibfnamefont {F.}~\bibnamefont
  {Corberi}}, \bibinfo {author} {\bibfnamefont {L.~F.}\ \bibnamefont
  {Cugliandolo}}, \bibinfo {author} {\bibfnamefont {F.}~\bibnamefont
  {Insalata}}, \ and\ \bibinfo {author} {\bibfnamefont {M.}~\bibnamefont
  {Picco}},\ }\href {\doibase 10.1103/PhysRevE.95.022101} {\bibfield  {journal}
  {\bibinfo  {journal} {Phys. Rev. E}\ }\textbf {\bibinfo {volume} {95}},\
  \bibinfo {pages} {022101} (\bibinfo {year} {2017})}\BibitemShut {NoStop}%
\bibitem [{\citenamefont {Corberi}\ \emph {et~al.}(2019)\citenamefont
  {Corberi}, \citenamefont {Cugliandolo}, \citenamefont {Insalata},\ and\
  \citenamefont {Picco}}]{CorCugInsPic19}%
  \BibitemOpen
  \bibfield  {author} {\bibinfo {author} {\bibfnamefont {F.}~\bibnamefont
  {Corberi}}, \bibinfo {author} {\bibfnamefont {L.~F.}\ \bibnamefont
  {Cugliandolo}}, \bibinfo {author} {\bibfnamefont {F.}~\bibnamefont
  {Insalata}}, \ and\ \bibinfo {author} {\bibfnamefont {M.}~\bibnamefont
  {Picco}},\ }\href {\doibase https://doi.org/10.1088/1742-5468/ab02ee}
  {\bibfield  {journal} {\bibinfo  {journal} {J. Stat. Mech.}\ }\textbf
  {\bibinfo {volume} {2019}},\ \bibinfo {pages} {043203} (\bibinfo {year}
  {2019})}\BibitemShut {NoStop}%
\bibitem [{\citenamefont {Agrawal}\ \emph {et~al.}(2022)\citenamefont
  {Agrawal}, \citenamefont {Corberi}, \citenamefont {Insalata},\ and\
  \citenamefont {Puri}}]{PhysRevE.105.034131}%
  \BibitemOpen
  \bibfield  {author} {\bibinfo {author} {\bibfnamefont {R.}~\bibnamefont
  {Agrawal}}, \bibinfo {author} {\bibfnamefont {F.}~\bibnamefont {Corberi}},
  \bibinfo {author} {\bibfnamefont {F.}~\bibnamefont {Insalata}}, \ and\
  \bibinfo {author} {\bibfnamefont {S.}~\bibnamefont {Puri}},\ }\href {\doibase
  10.1103/PhysRevE.105.034131} {\bibfield  {journal} {\bibinfo  {journal}
  {Phys. Rev. E}\ }\textbf {\bibinfo {volume} {105}},\ \bibinfo {pages}
  {034131} (\bibinfo {year} {2022})}\BibitemShut {NoStop}%
\bibitem [{\citenamefont {Duplantier}\ and\ \citenamefont
  {Saleur}(1988)}]{PhysRevLett.60.2343}%
  \BibitemOpen
  \bibfield  {author} {\bibinfo {author} {\bibfnamefont {B.}~\bibnamefont
  {Duplantier}}\ and\ \bibinfo {author} {\bibfnamefont {H.}~\bibnamefont
  {Saleur}},\ }\href {\doibase 10.1103/PhysRevLett.60.2343} {\bibfield
  {journal} {\bibinfo  {journal} {Phys. Rev. Lett.}\ }\textbf {\bibinfo
  {volume} {60}},\ \bibinfo {pages} {2343} (\bibinfo {year}
  {1988})}\BibitemShut {NoStop}%
\bibitem [{\citenamefont {Schramm}(2000)}]{schramm2000scaling}%
  \BibitemOpen
  \bibfield  {author} {\bibinfo {author} {\bibfnamefont {O.}~\bibnamefont
  {Schramm}},\ }\href@noop {} {\bibfield  {journal} {\bibinfo  {journal}
  {\href{https://doi.org/10.1007/BF02803524}{Isr. J. Math.}}\ }\textbf
  {\bibinfo {volume} {118}},\ \bibinfo {pages} {221} (\bibinfo {year}
  {2000})}\BibitemShut {NoStop}%
\bibitem [{\citenamefont {Cardy}(2005)}]{CARDY200581}%
  \BibitemOpen
  \bibfield  {author} {\bibinfo {author} {\bibfnamefont {J.}~\bibnamefont
  {Cardy}},\ }\href@noop {} {\bibfield  {journal} {\bibinfo  {journal}
  {\href{https://doi.org/10.1016/j.aop.2005.04.001}{Ann. Phys.}}\ }\textbf
  {\bibinfo {volume} {318}},\ \bibinfo {pages} {81} (\bibinfo {year}
  {2005})}\BibitemShut {NoStop}%
\bibitem [{\citenamefont {Smirnov}(2006)}]{Smirnov:2006cjh}%
  \BibitemOpen
  \bibfield  {author} {\bibinfo {author} {\bibfnamefont {S.}~\bibnamefont
  {Smirnov}},\ }\href@noop {} {\bibfield  {journal} {\bibinfo  {journal} {Proc.
  Int. Congr. Math.}\ }\textbf {\bibinfo {volume} {2}},\ \bibinfo {pages}
  {1421} (\bibinfo {year} {2006})}\BibitemShut {NoStop}%
\bibitem [{\citenamefont {Metropolis}\ and\ \citenamefont
  {Ulam}(1949)}]{metropolis1949monte}%
  \BibitemOpen
  \bibfield  {author} {\bibinfo {author} {\bibfnamefont {N.}~\bibnamefont
  {Metropolis}}\ and\ \bibinfo {author} {\bibfnamefont {S.}~\bibnamefont
  {Ulam}},\ }\href@noop {} {\bibfield  {journal} {\bibinfo  {journal}
  {\href{https://doi.org/10.2307/2280232}{J. Am. Stat. Assoc.}}\ }\textbf
  {\bibinfo {volume} {44}},\ \bibinfo {pages} {335} (\bibinfo {year}
  {1949})}\BibitemShut {NoStop}%
\bibitem [{\citenamefont {Agrawal}\ \emph {et~al.}(2021)\citenamefont
  {Agrawal}, \citenamefont {Corberi}, \citenamefont {Lippiello}, \citenamefont
  {Politi},\ and\ \citenamefont {Puri}}]{PhysRevE.103.012108}%
  \BibitemOpen
  \bibfield  {author} {\bibinfo {author} {\bibfnamefont {R.}~\bibnamefont
  {Agrawal}}, \bibinfo {author} {\bibfnamefont {F.}~\bibnamefont {Corberi}},
  \bibinfo {author} {\bibfnamefont {E.}~\bibnamefont {Lippiello}}, \bibinfo
  {author} {\bibfnamefont {P.}~\bibnamefont {Politi}}, \ and\ \bibinfo {author}
  {\bibfnamefont {S.}~\bibnamefont {Puri}},\ }\href {\doibase
  10.1103/PhysRevE.103.012108} {\bibfield  {journal} {\bibinfo  {journal}
  {Phys. Rev. E}\ }\textbf {\bibinfo {volume} {103}},\ \bibinfo {pages}
  {012108} (\bibinfo {year} {2021})}\BibitemShut {NoStop}%
\bibitem [{\citenamefont {Albano}\ \emph {et~al.}(2011)\citenamefont {Albano},
  \citenamefont {Bab}, \citenamefont {Baglietto}, \citenamefont {Borzi},
  \citenamefont {Grigera}, \citenamefont {Loscar}, \citenamefont
  {Rodr\'{i}guez}, \citenamefont {Puzzo},\ and\ \citenamefont
  {Saracco}}]{Albano_2011}%
  \BibitemOpen
  \bibfield  {author} {\bibinfo {author} {\bibfnamefont {E.~V.}\ \bibnamefont
  {Albano}}, \bibinfo {author} {\bibfnamefont {M.~A.}\ \bibnamefont {Bab}},
  \bibinfo {author} {\bibfnamefont {G.}~\bibnamefont {Baglietto}}, \bibinfo
  {author} {\bibfnamefont {R.~A.}\ \bibnamefont {Borzi}}, \bibinfo {author}
  {\bibfnamefont {T.~S.}\ \bibnamefont {Grigera}}, \bibinfo {author}
  {\bibfnamefont {E.~S.}\ \bibnamefont {Loscar}}, \bibinfo {author}
  {\bibfnamefont {D.~E.}\ \bibnamefont {Rodr\'{i}guez}}, \bibinfo {author}
  {\bibfnamefont {M.~L.~R.}\ \bibnamefont {Puzzo}}, \ and\ \bibinfo {author}
  {\bibfnamefont {G.~P.}\ \bibnamefont {Saracco}},\ }\href {\doibase
  10.1088/0034-4885/74/2/026501} {\bibfield  {journal} {\bibinfo  {journal}
  {Rep. Prog. Phys.}\ }\textbf {\bibinfo {volume} {74}},\ \bibinfo {pages}
  {026501} (\bibinfo {year} {2011})}\BibitemShut {NoStop}%
\bibitem [{\citenamefont {da~Silva}\ \emph {et~al.}(2002)\citenamefont
  {da~Silva}, \citenamefont {Alves},\ and\ \citenamefont
  {de~Fel{\i}cio}}]{DASILVA2002325}%
  \BibitemOpen
  \bibfield  {author} {\bibinfo {author} {\bibfnamefont {R.}~\bibnamefont
  {da~Silva}}, \bibinfo {author} {\bibfnamefont {N.~A.}\ \bibnamefont {Alves}},
  \ and\ \bibinfo {author} {\bibfnamefont {J.~D.}\ \bibnamefont
  {de~Fel{\i}cio}},\ }\href@noop {} {\bibfield  {journal} {\bibinfo  {journal}
  {Physics Letters A}\ }\textbf {\bibinfo {volume} {298}},\ \bibinfo {pages}
  {325} (\bibinfo {year} {2002})}\BibitemShut {NoStop}%
\bibitem [{\citenamefont {Duplantier}(2000)}]{PhysRevLett.84.1363}%
  \BibitemOpen
  \bibfield  {author} {\bibinfo {author} {\bibfnamefont {B.}~\bibnamefont
  {Duplantier}},\ }\href {\doibase 10.1103/PhysRevLett.84.1363} {\bibfield
  {journal} {\bibinfo  {journal} {Phys. Rev. Lett.}\ }\textbf {\bibinfo
  {volume} {84}},\ \bibinfo {pages} {1363} (\bibinfo {year}
  {2000})}\BibitemShut {NoStop}%
\bibitem [{\citenamefont {Wieland}\ and\ \citenamefont
  {Wilson}(2003)}]{PhysRevE.68.056101}%
  \BibitemOpen
  \bibfield  {author} {\bibinfo {author} {\bibfnamefont {B.}~\bibnamefont
  {Wieland}}\ and\ \bibinfo {author} {\bibfnamefont {D.~B.}\ \bibnamefont
  {Wilson}},\ }\href {\doibase 10.1103/PhysRevE.68.056101} {\bibfield
  {journal} {\bibinfo  {journal} {Phys. Rev. E}\ }\textbf {\bibinfo {volume}
  {68}},\ \bibinfo {pages} {056101} (\bibinfo {year} {2003})}\BibitemShut
  {NoStop}%
\bibitem [{\citenamefont {Lawler}\ \emph {et~al.}(2004)\citenamefont {Lawler},
  \citenamefont {Schramm},\ and\ \citenamefont {Werner}}]{10.2307/3481623}%
  \BibitemOpen
  \bibfield  {author} {\bibinfo {author} {\bibfnamefont {G.~F.}\ \bibnamefont
  {Lawler}}, \bibinfo {author} {\bibfnamefont {O.}~\bibnamefont {Schramm}}, \
  and\ \bibinfo {author} {\bibfnamefont {W.}~\bibnamefont {Werner}},\ }\href
  {http://www.jstor.org/stable/3481623} {\bibfield  {journal} {\bibinfo
  {journal} {The Annals of Probability}\ }\textbf {\bibinfo {volume} {32}},\
  \bibinfo {pages} {939} (\bibinfo {year} {2004})}\BibitemShut {NoStop}%
\bibitem [{\citenamefont {Smirnov}(2010)}]{smirnov2010conformal}%
  \BibitemOpen
  \bibfield  {author} {\bibinfo {author} {\bibfnamefont {S.}~\bibnamefont
  {Smirnov}},\ }\href@noop {} {\bibfield  {journal} {\bibinfo  {journal}
  {Annals of mathematics}\ }\textbf {\bibinfo {volume} {172}},\ \bibinfo
  {pages} {1435} (\bibinfo {year} {2010})}\BibitemShut {NoStop}%
\bibitem [{\citenamefont {Smirnov}(2001)}]{SMIRNOV2001239}%
  \BibitemOpen
  \bibfield  {author} {\bibinfo {author} {\bibfnamefont {S.}~\bibnamefont
  {Smirnov}},\ }\href {\doibase https://doi.org/10.1016/S0764-4442(01)01991-7}
  {\bibfield  {journal} {\bibinfo  {journal} {Comptes Rendus de l'Academie des
  Sciences - Series I - Mathematics}\ }\textbf {\bibinfo {volume} {333}},\
  \bibinfo {pages} {239} (\bibinfo {year} {2001})}\BibitemShut {NoStop}%
\bibitem [{\citenamefont {Janssen}\ \emph {et~al.}(1995)\citenamefont
  {Janssen}, \citenamefont {Oerding},\ and\ \citenamefont
  {Sengespeick}}]{Janssen_1995}%
  \BibitemOpen
  \bibfield  {author} {\bibinfo {author} {\bibfnamefont {H.~K.}\ \bibnamefont
  {Janssen}}, \bibinfo {author} {\bibfnamefont {K.}~\bibnamefont {Oerding}}, \
  and\ \bibinfo {author} {\bibfnamefont {E.}~\bibnamefont {Sengespeick}},\
  }\href {\doibase 10.1088/0305-4470/28/21/012} {\bibfield  {journal} {\bibinfo
   {journal} {J. Phys. A: Math. Gen.}\ }\textbf {\bibinfo {volume} {28}},\
  \bibinfo {pages} {6073} (\bibinfo {year} {1995})}\BibitemShut {NoStop}%
\bibitem [{\citenamefont {Heuer}(1993)}]{Heuer_1993}%
  \BibitemOpen
  \bibfield  {author} {\bibinfo {author} {\bibfnamefont {H.-O.}\ \bibnamefont
  {Heuer}},\ }\href {\doibase 10.1088/0305-4470/26/6/007} {\bibfield  {journal}
  {\bibinfo  {journal} {J. Phys. A: Math. Gen.}\ }\textbf {\bibinfo {volume}
  {26}},\ \bibinfo {pages} {L333} (\bibinfo {year} {1993})}\BibitemShut
  {NoStop}%
\bibitem [{\citenamefont {da~Silva}\ \emph {et~al.}(2009)\citenamefont
  {da~Silva}, \citenamefont {Fulco},\ and\ \citenamefont {Nobre}}]{Silva_2009}%
  \BibitemOpen
  \bibfield  {author} {\bibinfo {author} {\bibfnamefont {L.~F.}\ \bibnamefont
  {da~Silva}}, \bibinfo {author} {\bibfnamefont {U.~L.}\ \bibnamefont {Fulco}},
  \ and\ \bibinfo {author} {\bibfnamefont {F.~D.}\ \bibnamefont {Nobre}},\
  }\href {\doibase 10.1088/0953-8984/21/34/346005} {\bibfield  {journal}
  {\bibinfo  {journal} {J. Phys.: Condens. Matter}\ }\textbf {\bibinfo {volume}
  {21}},\ \bibinfo {pages} {346005} (\bibinfo {year} {2009})}\BibitemShut
  {NoStop}%
\bibitem [{Note2()}]{Note2}%
  \BibitemOpen
  \bibinfo {note} {The pinning time $t_p$ is much smaller than the relaxation
  time $t_{\protect \rm eq} \simeq L^{z_c}$, and in the thermodynamic limit,
  $t_p/t_{\protect \rm eq} \rightarrow 0$. See Ref.~\cite {BlaCorCugPic14} for
  details.}\BibitemShut {Stop}%
\end{thebibliography}%

\end{document}